\documentclass[12pt]{iopart}
\usepackage{times}
\usepackage{graphicx}

\def\red#1{{
#1
}}

\def\beq{\begin{equation}}
\def\eeq{\end{equation}} 
\def\beqn{\begin{eqnarray}}
\def\eeqn{\end{eqnarray}}

\def\AEF{Faraggi A E }
\def\NPB#1#2#3{ #2  {\it Nucl.\ Phys.\ B}\/ {\bf #1}  #3}
\def\PLB#1#2#3{ #2  {\it Phys.\ Lett.\ B}\/ {\bf #1}  #3}
\def\PLA#1#2#3{ #2  {\it Phys.\ Lett.\ A}\/ {\bf #1}  #3}
\def\PRD#1#2#3{ #2  {\it Phys.\ Rev.\ D}\/ {\bf #1}  #3}

\def\PRT#1#2#3{  #2  {\it Phys.\ Rep.}\/ {\bf#1}  #3}
\def\MODA#1#2#3{ #2  {\it Mod.\ Phys.\ Lett.\ A}\/ {\bf #1}  #3}
\def\IJMP#1#2#3{ #2  {\it Int.\ J.\ Mod.\ Phys.\ A}\/ {\bf #1} #3}
\def\nuvc#1#2#3{ #2  {\it Nuovo Cimento}\/ {\bf #1}{\it C} #3}
\def\APJ#1#2#3{  #2  {\it Astrophys.\ J.}\/ {\bf #1} #3}
\def\JHEP#1#2#3{ #2  {\it JHEP } {\bf #1} #3} 
\font\bigbf=cmssbx10 scaled\magstep2

\usepackage{latexsym}
\input epsf
\begin{document}
\rightline{OUTP--02--34P}

\title{Phenomenological aspects of M--theory}
\author{Alon E. Faraggi\footnote{Invited talk presented 
at Beyond the Desert 02, Oulu, Finland 2-7 June 2002.}
}
\address{Theoretical Physics Department, University of Oxford,
Oxford OX1 3NP, UK\\ 
and Theory Division, CERN, CH--1211 Geneva, Switzerland}
\begin{abstract}
The Standard Model data suggests the realization of grand unification
structures in nature, in particular that of $SO(10)$.
A class of string vacua that preserve the $SO(10)$ embedding,
are the three generation free fermion heterotic--string models,
that are related to $Z_2\times Z_2$ orbifold compactification.
Attempts to use the M--theory framework to explore further this
class of models are discussed. Wilson line breaking of the
$SO(10)$ GUT symmetry results in super--heavy meta--stable
states, which produce several exotic dark matter and UHECR
candidates, with differing phenomenological characteristics.
Attempts to develop the tools to decipher the properties
of these states in forthcoming UHECR experiments are discussed.
It is proposed that quantum mechanics follows from an equivalence
postulate, that may lay the foundations for the rigorous
formulation of quantum gravity.

\end{abstract}

\section{Introduction}

\begin{figure}[t]
\centerline{\epsfxsize 3.0 truein \epsfbox {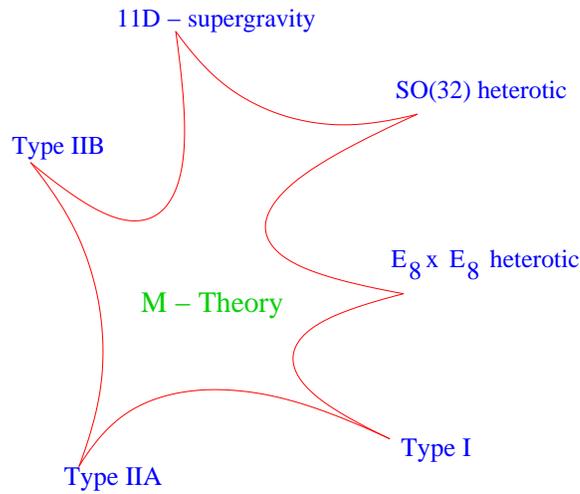}}
\caption{M--theory picture of string theory}  
\label{mtheory}
\end{figure}

Over the past few years important progress has been achieved in the basic 
understanding of string theory \cite{Mtheoryreviews}.
The picture that emerged, and which 
is depicted qualitatively in fig. \ref{mtheory}, is that the different
string theories in ten dimensions are perturbative limits of a single
more fundamental theory. The vital question remains,
how to relate these advances to observational data, as seen
in terrestrial and astrophysical experiments. 
On the other hand, particle physics experiments over the
past century cumulated in the Standard Particle Model.
This model, and more importantly its experimental verification,
represent the pinnacle of scientific achievement,
and a source of elation for every participant in
this magnificent endeavor. 

\begin{figure}[t]
\centerline{\epsfxsize 3.0 truein \epsfbox {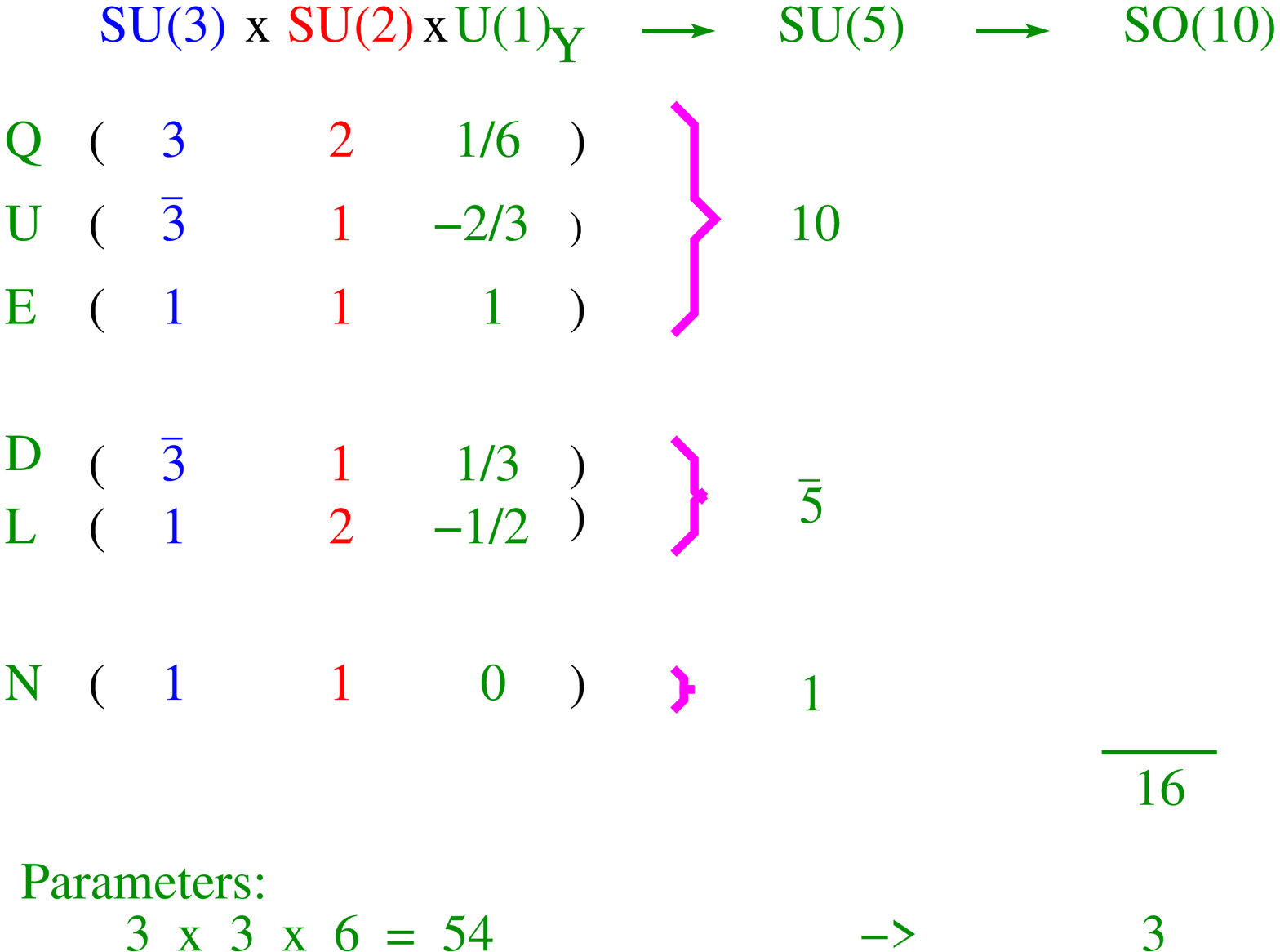}}
\caption{Standard Model multiplet structure}  
\label{smmultiplets}
\end{figure}

What lesson should we extract then from the brilliant success of the Standard
Model. An important aspect of the Standard Model is
its multiplet structure, which is
exhibited in fig. \ref{smmultiplets}.
The remarkable feature of the Standard Model
is the embedding of its gauge and matter states into representations of
Grand Unified Theories (GUTs) \cite{gg}. This astonishing 
fact seems almost a ``triviality''. Namely,
from a mathematical perspective the representations and associated
group theory are rather mundane. The Standard Model
multiplet structure is, however,
what we observe experimentally, in multi--billion
dollar collider experiments, and
a priori there was no particular reason for this
structure to emerge. The GUT embedding is most economical in
the framework of $SO(10)$ \cite{so10}, in which all the Standard Model
states, including the right--handed neutrinos,
which are desirable for neutrino masses and
oscillations, in each generation, are embedded
in a single 16 representation. Strikingly, each of the three generations
fit into a spinorial 16
representation of $SO(10)$. If we regard the 
quantum numbers of the Standard Model states as
experimental observables, as they were in the process of its
experimental discovery, then, prior to unification,
we count 54 free parameters. These include the quantum numbers
of the three generations, each composed of six separate multiplets,
under the Standard Model three group factors.
The embedding in $SO(10)$ therefore reduces this number
of free parameters from 54 to 3, which is the number of
$SO(10)$ 16 representations needed to accommodate the
Standard Model states. This to me looks like a true
wonder of nature, and should serve as
the guide in attempts to understand the fundamental 
origins of the Standard Model. 

While the Standard Model is confirmed, 
and its GUT generalizations are well motivated, by the experimental data,
they leave many issues unresolved. The Higgs sector
is unresolved experimentally and the associated electroweak
symmetry breaking mechanism is still shrouded in mystery. 
Furthermore, the masses of the Higgs scalars cannot be protected
against radiative corrections from the fundamental cut--off
scale, a severe  hierarchy problem.
Finally the replication of the family flavors and the mass spectrum
are not explained
in the framework of Grand Unified Theories. It would therefore 
appear that these additional structures arise from physics
above the GUT scale, where the gravitational interaction play
a predominant role. String theory provides the unique framework
to study how Planck scale physics determine these Standard Model
structures. 

\subsection{High or low?}

Indeed string theory, in its
several perturbative incarnations, inspires much of the 
contemporary studies of physics beyond the 
Standard Model. From the experimental perspective the 
burning issue is the nature of the electroweak symmetry
breaking mechanism and of the Higgs state. From a theoretical
perspective the suppression of the electroweak scale, 
as compared to the fundamental GUT and gravitational scales,
raises basic questions. To solve this puzzle two major 
school of thoughts have been developed. The first assumes 
the existence of new strongly interacting sector at the 
electroweak symmetry breaking scale. These include the old technicolor
and composite Higgs theories, and their modern incarnations, 
in the form of large extra dimensions and warped compactifications. 
The second assumes that the Standard Model remains in the perturbative
regime up to a scale which is of the order of the GUT and Planck
scales. To stabilize the electroweak scale against radiative corrections
from the larger scales one has to evoke the existence of a new symmetry,
supersymmetry. The possibility of having a new strongly interacting 
sector at the TeV scale received a strong impetus in recent years
due to the realization that the fundamental string unification
scale can be lowered in the framework of M-theory \cite{lowerscale}.

The problem that supersymmetry or a new strongly interacting sector 
at the TeV scale therefore address is the suppression of the 
scalar sector masses as compared to the fundamental unification scale,
which is known as ``The hierarchy problem''.
The solutions invoked involve either the cancellation between 
fermionic and bosonic contributions to the Higgs boson masses
(supersymmetry) or imposing that the cut--off
scale is at the TeV scale (new strongly interacting sector;
large extra dimensions; warped compactifications; etc).
The natural question is whether these are mere alternatives, or
whether one choice is preferred over the other.
To contemplate this question we should examine the experimental data.
The high scale unification paradigm is motivated by the Standard Model and
Grand Unified Theories. In this context the hierarchy between the
electroweak scale and the GUT or Planck scales arises due to the
logarithmic running of the Standard Model parameters \cite{gqw}.
In the gauge sector of the Standard Model this logarithmic running
has in fact been confirmed experimentally in the energy regime 
which is accessible to collider experiments. Namely by measuring
the evolution of the Standard Model gauge parameters between
the QCD scale and the ~200 GeV scale probed at LEP, the logarithmic 
evolution agrees with observations. The assumption of the high
scale unification is also in qualitative agreement with the 
data in the gauge and matter sectors of the Standard Model,
as is exemplified by the resulting predictions for
$\sin^2\theta_W(M_Z)$ \cite{gqw,gcumssm} and for the mass ratio
$M_b(M_Z)/M_\tau(M_Z)$ \cite{mbmtau}.

Thus supersymmetry maintains the logarithmic running in the gauge and
matter sectors and restores it in the scalar sector. On the 
other hand a new strongly interacting sector
(large extra dimensions; warped compactifications; etc)
at the TeV scale annuls the logarithmic running in the
gauge and matter sectors.
We therefore note that, while supersymmetry has not yet been 
observed experimentally, logarithmic running, which is the
origin of the hierarchy problem, has been observed experimentally
in the gauge and matter sectors of the Standard Model.
Therefore, theories which preserve this logarithmic running,
which is the root for the hierarchical scales, are 
superior to those that do not. 

\section{Pheno--M--enology}\label{mpheno}

The Standard Model and the low energy observed data support
the existence of high scale unification and the big desert 
scenario. On the other hand, understanding many of the 
properties of the low energy data necessitates the incorporation
of gravity into the unification program. Superstring theory
provides a consistent perturbative approach to quantum gravity
in which many of these issues can be studied. Due to the fundamental
advances of recent years we know that the different string
theories in ten dimensions are perturbative limits of 
a more fundamental theory, traditionally dubbed M--theory.
This picture is qualitatively depicted in figure \ref{mtheory}.
In this context, the true fundamental theory of nature should have some
nonperturbative realization. However, at present all we know
about this more basic theory, are its perturbative string limits.
Therefore, we should regard the perturbative string limits as
providing tools to probe the properties of the fundamental nonperturbative
vacuum in the different perturbative limits.
As posited above the remarkable property of the Standard Model
spectrum is the embedding of the chiral generations in the spinorial
16 representation of $SO(10)$. Thus, we may demand the existence of 
a viable perturbative string limit which preserve this embedding.
The only perturbative string limit which enables the $SO(10)$
embedding of the Standard Model spectrum is the heterotic $E_8\times E_8$
string. The reason being that only this limit produces the 
spinorial 16 representation in the perturbative massless spectrum. 
Therefore, if we would like to preserve the $SO(10)$ embedding of the 
Standard Model spectrum, the M--theory limit which we should use is
the perturbative heterotic string. In this respect it may well
be that other perturbative string limits may provide more useful
means to study other properties of the true nonperturbative vacuum,
such as dilaton and moduli stabilization. This suggests the following 
approach to exploration of M--theory phenomenology. Namely, the 
true M-theory vacuum has some nonperturbative realization that at
present we do not know how to formulate. This vacuum is at finite
coupling and is located somewhere in the space of vacua enclosed
by figure \ref{mtheory}. The properties of the true vacuum can
however be probed in the perturbative string limits. We may
hypothesize that in any of these limits one still needs to compactify
to four dimensions. Namely, that the true M--theory vacuum
can still be formulated with four non--compact and all the other
dimensions are compact. Then, suppose that in some of the limits
we are able to identify a specific class of compactifications
that possess appealing phenomenological properties. The new 
M--theory picture suggests that we can then explore the possible 
properties of the M--theory vacuum by studying compactifications
of the other perturbative string limits on the same class of
compactifications. This is the approach that we try to develop
in the work described here.

The first task is therefore to construct string vacua that are as
realistic as possible. Such a model of unification should of course
satisfy a large number of constraints, a few of which are listed below,

\smallskip
\centerline{{$\underline{{\hbox{~~~~~~~~~~~~~~~~~~~~~~~~~~~~~~~~~~~~~}}}$}}
{}~~~~~$1.$ Gauge group ~$\longrightarrow$~$SU(3)\times SU(2)\times
U(1)_Y$
~~~~~~~~~~~~~~~{$U(1)\in SO(10)$}

{}~~~~~$2.$ Contains three generations

{}~~~~~$3.$ Proton stable ~~~~~~~~($\tau_{\rm P}>10^{30+}$ years)

{}~~~~~$4.$ N=1 supersymmetry~~~~~~~~(or N=0)

{}~~~~~$5.$ Contains Higgs doublets $\oplus$ potentially realistic
Yukawa couplings

{}~~~~~$6.$ Agreement with $\underline{\sin^2\theta_W}$ and
$\underline{\alpha_s}$ at $M_Z$ (+ other observables).

{}~~~~~$7.$ Light left--handed neutrinos

{}~~~~~~~~~~~~~$~8.$ $SU(2)\times U(1)$ breaking

{}~~~~~~~~~~~~~$~9.$ SUSY breaking

{}~~~~~~~~~~~~~$10.$ No flavor changing neutral currents

{}~~~~~~~~~~~~~$11.$ No strong CP violation

{}~~~~~~~~~~~~~$12.$ Exist family mixing and weak CP violation

{}~~~~~$13.$ +~~ {\bf ...}

\red{{}~~~~~$14.$ +~~~~~~~~~~~~~~~~{\bigbf{NO FREE EXOTICS}}}

\centerline{{$\underline{{\hbox{~~~~~~~~~~~~~~~~~~~~~~~~~~~~~~~~~~~~~}}}$}}
\smallskip

The embedding of the Standard Model spectrum in $SO(10)$ representations
implies that the weak hypercharge should have the canonical $SO(10)$
embedding. The $SO(10)$ symmetry need not be realized in
an effective
field theory but can be broken directly at the string level.
In which case the Standard Model spectrum still
arises from $SO(10)$ representations, but the $SO(10)$
non--Abelian states, beyond the Standard Model, 
are projected out by the GSO projections. However, 
if we take the Standard Model $SO(10)$ embedding
as a necessary requirement this means that the
weak hypercharge must have the standard $SO(10)$
embedding with $k_Y=5/3$.

The construction of realistic superstring vacua proceeds by studying
compactification of the heterotic string from ten to four dimensions. 
Various methods can be used for this purpose which include
geometric and algebraic tools, and each has its advantages and
disadvantages. One class of models utilizes compactifications
on Calabi--Yau 3--folds that give rise to an $E_6$ observable
gauge symmetry, which is broken further by Wilson lines
to $SU(3)^3$ \cite{suthree}. This type of geometrical 
compactifications correspond at special points to
conformal theories which have $(2,2)$ world--sheet supersymmetry.
Similar compactifications which have only (2,0) world--sheet
supersymmetry have also been studied and can lead to compactifications
with $SO(10)$ and $SU(5)$ observable gauge groups \cite{twozero}.
The analysis of this type of compactification
is complicated due to the fact that they do not correspond
to free world--sheet theories. Therefore, it is difficult
to calculate the parameters of the Standard Model in these constructions.
On the other hand they provide a sophisticated mathematical
window to the underlying geometry.

The next class of superstring vacua are the orbifold models \cite{dhvw}.
Here one starts with a compactification
of the heterotic string on a flat torus,
using the Narain prescription \cite{narain},
and utilizes free world--sheet bosons. The Narain
lattice is moded out by some discrete symmetries which are the
orbifold twisting. An important class of models of this type are
the $Z_3$ orbifold models \cite{zthree}. These give rise to
three generation models with $SU(3)\times SU(2)\times U(1)^n$
gauge group. A deficiency of this class of models is that
they do not give rise to the standard $SO(10)$ embedding of
the Standard Model spectrum. Consequently,
the normalization of $U(1)_Y$, relative to the non--Abelian
currents, is typically larger than 5/3, the standard $SO(10)$ normalization.
This results generically in disagreement with the observed low energy
values for $\sin^2\theta_W(M_Z)$ and $\alpha_s(M_Z)$. 
A thorough analysis of $Z_3$ orbifold model was performed
recently by Giedt \cite{giedt},
who also demonstrated the existence of models in this class which
admit the $SU(5)$ embedding of the Standard Model gauge group.

The final class of perturbative string compactifications
consist of those that are constructed from world--sheet
conformal field theories. The simplest of those correspond
to the free fermionic formulation in which all the degrees
of freedom needed to cancel the conformal anomaly are represented
in terms of free fermions propagating on the string world--sheet
\cite{FFF}. This in turn correspond to compactifications at fixed radii, 
and non--trivial background metric and antisymmetric tensor fields.
More complicated world--sheet conformal field theories can
also be formulated and, in general, correspond to compactifications
at fixed radii \cite{gepner}. 

It is important to emphasize that we expect that the different
formulations of string compactifications are related.
This in particular means that by utilizing these various tools
one is not probing different physics. Thus, the different constructions
can be used to study different aspects of specific classes of
string compactifications. Naturally, we expect that some issues
are better examined by using a particular formulation, whereas
for others another may prove advantageous. 

The concrete class of string models that form the basis 
of our studies here are those that are constructed in 
the free fermionic formulation. In the next section I 
briefly review the models.
It is vital to note that this class of string compactifications
correspond to $Z_2\times Z_2$ orbifold twisting of the Narain 
lattice which is realized at the free fermionic point in moduli space,
and augmented with additional Wilson lines. The correspondence
of a subset of these models with the $Z_2\times Z_2$ orbifold
compactification \cite{foc}
is exploited in attempts 
to elevate the study of these models to the nonperturbative regime. 
This construction gives rise naturally to two of the
key propertied of the observed Standard Model particle spectrum.
Specifically, it naturally produces three chiral generations
with the canonical $SO(10)$ embedding. Roughly speaking
the origin of three generations arises in the $Z_2\times Z_2$
orbifold based constructions, due to the fact that we are
dividing a six dimensional compactified manifold into factors
of two. Specifically, we have that the $Z_2\times Z_2$ orbifold
twisting of a six dimensional compactified lattice produces
exactly three twisted sectors. In the realistic free fermionic
models one generation is obtained from each of the three
twisted sectors. Thus, heuristically speaking the origin of
three generations in these models is traced to
$${6\over 2}=3.$$ The natural question in the context
of M--theory is how this naive understanding extend
to the nonperturbative domain. However, the mere fact that
we trace here the origin of three generations to the basic 
structure of the underlying compactification suggests that
it is independent of the perturbative expansion.
The second crucial property of the realistic free fermionic models
is that they preserve the canonical $SO(10)$ embedding
of the Standard Model spectrum. Consequently, the free
fermionic construction produces a large class of three
generation models, with differing phenomenological
details. 

\section{Free fermionic models}\label{ffmodels}
I discuss here briefly the structure of the free fermionic 
models that forms the basis of our explorations. 
Phenomenological aspects of these models have been reviewed
in the past in this conference series \cite{btd9799}.
I recap the structure of the models which is 
most relevant for the specific pheno--M--enological aspects
of M--theory that are of interest here. 

A model in the free fermionic
formulation \cite{FFF} is defined by a set of  boundary
condition basis vectors, and one--loop GSO phases, which are
constrained by the string consistency requirements,
and completely determine the vacuum structure of the models.
The physical spectrum is obtained by applying the generalized 
GSO projections. The Yukawa couplings 
and higher order nonrenormalizable terms in the superpotential
are obtained by calculating correlators between vertex
operators \cite{KLN}. 
The realistic free fermionic
models produce an ``anomalous'' $U(1)$ symmetry,
which generates a Fayet--Iliopoulos D--term \cite{dsw},
and breaks supersymmetry at the Planck scale. Supersymmetry
is restored by assigning non vanishing VEVs to a set of Standard
Model singlets in the massless string spectrum along flat F and D
directions. In this process nonrenormalizable terms,
$\langle{V_1^fV_2^fV_3^b\cdot\cdot\cdot\cdot V_N^b\rangle},$
become renormalizable operators, 
$V_1^fV_2^fV_3^b{{\langle V_4^b\cdots V_N^b\rangle}/{M^{N-3}}}$
in the effective low energy field theory.

The first five basis vectors of the realistic free fermionic
models consist of the NAHE set \cite{nahe}.
The gauge group after the NAHE set is
$SO(10)\times E_8\times SO(6)^3$ with $N=1$ space--time supersymmetry, 
and 48 spinorial $16$ of $SO(10)$, sixteen from each sector $b_1$,
$b_2$ and $b_3$. The three sectors $b_1$, $b_2$ and $b_3$ are
the three twisted sectors of the corresponding $Z_2\times Z_2$
orbifold compactification. The $Z_2\times Z_2$ orbifold is special
precisely because of the existence of three twisted sectors,
with a permutation symmetry with respect to the horizontal $SO(6)^3$
charges. 

The NAHE set is common to a large class of three generation
free fermionic models. The construction proceeds by adding to the
NAHE set three additional boundary condition basis vectors
which break $SO(10)$ to one of its subgroups: $SU(5)\times U(1)$
\cite{revamp}, $SO(6)\times SO(4)$ \cite{patisalamstrings},
$SU(3)\times SU(2)\times U(1)^2$ \cite{fny,eu,top,cfn},
or $SU(3)\times U(1)\times SO(4)$ \cite{lrsstringmodels}.
At the same time the number of generations is reduced to
three, one from each of the sectors $b_1$, $b_2$ and $b_3$.
The various three generation models differ in their
detailed phenomenological properties. However, many of
their characteristics can be traced back to the underlying
NAHE set structure. One such important property to note
is the fact that as the generations are obtained
from the three twisted sectors $b_1$, $b_2$ and $b_3$,
they automatically possess the Standard $SO(10)$ embedding.
Consequently the weak hypercharge, which arises as
the usual combination $U(1)_Y=1/2 U(1)_{B-L}+ U(1)_{T_{3_R}}$,
has the standard $SO(10)$ embedding.

The massless spectrum of the realistic free fermionic models
then generically contains three generations from the
three twisted sectors $b_1$, $b_2$ and $b_3$, which are
charged under the horizontal symmetries. The Higgs spectrum
consists of three pairs of electroweak doublets from the 
Neveu--Schwarz sector plus possibly additional one or
two pairs from a combination of the two basis vectors
which extend the NAHE set. Additionally the models 
contain a number of $SO(10)$ singlets which are
charged under the horizontal symmetries and
a number of exotic states. 

Exotic states
arise from the basis vectors which extend the NAHE
set and break the $SO(10)$ symmetry \cite{ccf}. Consequently, they
carry either fractional $U(1)_Y$ or $U(1)_{Z^\prime}$ charge.
Such states are generic in superstring models
and impose severe constraints on their validity.
In some cases the exotic fractionally charged
states cannot decouple from the massless
spectrum, and their presence invalidates otherwise
viable models \cite{otherrsm,penn}.
In the NAHE based models the fractionally
charged states always appear in vector--like
representations. Therefore, in general mass
terms are generated from renormalizable or nonrenormalizable
terms in the superpotential. However, the mass terms which arise
from non--renormalizable terms will in general be suppressed,
in which case the fractionally charged states may have
intermediate scale masses.
The analysis of ref. \cite{cfn} demonstrated the
existence of free fermionic models with solely the
MSSM spectrum in the low energy effective field theory of the
Standard Model charged matter. 
In general, unlike the ``standard'' spectrum, the ``exotic'' spectrum is
highly model dependent. 

\section{Phenomenological studies of free fermionic string models}

I summarize here some of the highlights of the phenomenological studies
of the free fermionic models. This
demonstrates that the free fermionic string models indeed
provide the arena for exploring many the questions relevant
for the phenomenology of the Standard Model and Unification. 
The lesson that should be extracted is that
the underlying structure of these models, generated
by the NAHE set, produces the right features for
obtaining realistic phenomenology. It provides further
evidence for the assertion that the true string
vacuum is connected to the $Z_2\times Z_2$ orbifold
in the vicinity of the free fermionic point in the
Narain moduli space. Many
of the important issues relating to the phenomenology of
the Standard Model and supersymmetric unification have been
discussed in the past in several prototype free fermionic
heterotic string models. These studies have been reviewed in
the past and I refer to the original literature and additional
review references \cite{review,btd9799}.
These include among others: top quark mass prediction \cite{top}, several
years prior to the actual observation by the CDF/D0 collaborations
\cite{cdfd0};
generations mass hierarchy \cite{NRT}; CKM mixing \cite{CKM};
superstring see--saw mechanism \cite{seesaw}; Gauge coupling
unification \cite{gcu}; Proton stability \cite{ps}; and
supersymmetry breaking and squark degeneracy \cite{fp2}.
Additionally,
it was demonstrated in ref. \cite{cfn} that at low energies the model
of ref. \cite{fny}, which may be viewed as a
prototype example of a realistic free fermionic model,
produces in the observable sector solely the MSSM charged spectrum.  
Therefore, the model of ref. \cite{fny}, supplemented
with the flat F and D solutions of ref. \cite{cfn}, provides
the first examples in the literature of a string model
with solely the MSSM charged spectrum
below the string scale. Thus, for the first time it provides
an example of a long--sought Minimal Superstring Standard Model!
We have therefore identified
a neighborhood in string moduli space which is potentially
relevant for low energy phenomenology. While we can suggest
arguments, based on target--space duality considerations why this
neighborhood may be selected, we cannot credibly argue that
similar results cannot be obtained in other regions
of the string moduli space. Nevertheless, the results summarized
here provide the justification for further
explorations of the free fermionic models. Furthermore,
they provide motivation to study these models
in the nonperturbative context of M--theory. 
In this context the basis for our studies is the 
connection of the free fermionic models with the 
$Z_2\times Z_2$ orbifold, to which I turn in section \ref{z2z2orbifold}.

I would like to emphasize that it is not suggested that any of the
realistic free fermionic models is the true vacuum of our world.
Indeed such a claim would be folly. Each of the phenomenological
free fermionic models has its shortcomings, that if time and space
would have allowed could have been detailed. While in principle
the phenomenology of each of these models may be improved
by further detailed analysis of supersymmetric flat directions,
it is not necessarily the most interesting avenue for exploration.
The aim of the studies outlined above is to demonstrate that
all of the major issues, pertaining to the phenomenology of the 
Standard Model and unification, can in principle be
addressed in the framework of the free fermionic models,
rather than to find the explicit solution that accommodates
all of these requirements simultaneously. The reason being
that even within this space of solutions there is till a vast
number of possibilities, and we lack the guide to select the
most promising one. What is being proposed is that these
phenomenological studies suggest that the true string vacuum
may share some of the gross structure of the free fermionic models.
Namely, it will possess the structure of the $Z_2\times Z_2$
orbifold in the vicinity of the free fermionic point in the
Narain moduli space. This perspective provides the motivation
for the continued interest in the detailed study of this gross structure,
and specifically in the framework of M--theory, as discussed below.

\section{Toward string predictions}\label{towards}

Before turning to the recent M--theory related studies,
I briefly discuss possible signatures of the string
models, beyond the Standard Model.
Following the demonstration of phenomenological viability of
a class of string compactifications, it is sensible
to seek possible experimental signals that
may provide evidence for specific models in particular,
and for string theory in general. This is in essence a secondary task
as the first duty is to reproduce the observed physics
of the Standard Model. 
With these priorities in mind there are several possible
exotic signatures that have been discussed in the past. 
These include the possibility of extra $U(1)$'s \cite{zp};
specific supersymmetric spectrum scenarios \cite{dedes}; R--parity
violation \cite{rp}; and exotic matter \cite{ccf}. 
R--parity violation at an observable rate is
an intriguing but somewhat remote possibility.
The problem is that in string models
if R--parity is violated at the same time one
expects to generate fast proton decay.
The model of ref. \cite{custodial} provides an example how
R--parity violation can arise in superstring theory.
This string model gives rise to custodial symmetries
which allow lepton number violation
while forbidding baryon number violation.

The second possibility is that of additional $Z^\prime$ gauge bosons
in energy regimes that may be accessible to future colliders
\cite{zp,thor}.
This possibility is motivated by the need to suppress proton decay,
while at the same time insuring that the masses of the left--handed
neutrinos are adequately suppressed \cite{jogesh,thor}.
In the context of string constructions
a conflict may arise due to the fact that one typically utilizes VEVs
that break the $B-L$ symmetry spontaneously. Such VEVs are used, for 
example, in the flipped $SU(5)$ to break the $SU(5)$ GUT symmetry.
In other string models such VEVs typically must be used to generate
a seesaw mechanism. The need to use $B-L$ breaking VEVs for this purpose
follows from the absence of the $SO(10)$ 126 representation
in the perturbative massless string spectrum \cite{dienes}. On the other
hand the $B-L$ breaking VEVs may induce rapid proton decay 
from dimension four operators. However, dimension four
and five proton decay mediating operators may be adequately
suppressed if the dangerous terms are forbidden by an additional
$U(1)$ symmetry. In this case the effective magnitude of
the proton decay mediating operators is determined
by the symmetry breaking scale of the extra $U(1)$,
and may be adequately suppressed provided that $\Lambda_{Z^\prime}$
is sufficiently small. String models
produce $U(1)$ symmetries with the required properties that are
external to the GUT symmetries. Typically, they will be family dependent
and are therefore also constrained by the flavor data. However,
such a $U(1)$ symmetry can remain unbroken down to $\sim30$TeV and may
have observable consequences in future colliders and/or flavor experiments.

The last possibility that I discuss here is that of exotic
matter, which arises in superstring models
because of the breaking of the non--Abelian symmetries
by Wilson--lines \cite{fccp,ccf}. It is therefore a unique
signature of superstring unification, 
which does not arise in field theory GUTs.
While the existence of such states
imposes severe constraints on otherwise valid
string models \cite{otherrsm}, provided that
the exotic states are either confined or sufficiently heavy,
they can give rise to exotic signatures.
For example, they can produce heavy dark
matter candidates, possibly with observable
consequences \cite{elnfc,fop,cfp}. In section
\ref{uhecr} I will elaborate on the exotic matter
in the string models and possible observable consequences.

\section{$Z_2\times Z_2$ orbifold correspondence}\label{z2z2orbifold}

The key property of the fermionic models that is exploited
in trying to elevate the analysis of these models to the 
nonperturbative domain of M--theory is the correspondence
with the $Z_2\times Z_2$ orbifold compactification. 
As discussed in section  \ref{ffmodels} the construction of the
realistic free fermionic models can be divided into two
parts. The first part consist of the NAHE--set basis vectors
and the second consists of the additional boundary conditions,
$\{\alpha,\beta,\gamma\}$ that correspond to Wilson lines
in the orbifold language.
The correspondence of the NAHE-based free fermionic models  
with the orbifold construction is illustrated
by extending the NAHE set, $\{ 1,S,b_1,b_2,b_3\}$, by one additional   
boundary condition basis vector \cite{foc},
\beq
\xi_1=(0,\cdots,0\vert{\underbrace{1,\cdots,1}_{{\bar\psi^{1,\cdots,5}},
{\bar\eta^{1,2,3}}}},0,\cdots,0)~.
\label{vectorx}
\eeq
With a suitable choice of the GSO projection coefficients the
model possesses an ${\rm SO}(4)^3\times {\rm E}_6\times {\rm U}(1)^2
\times {\rm E}_8$ gauge group
and $N=1$ space-time supersymmetry. The matter fields
include 24 generations in the 27 representation of
${\rm E}_6$, eight from each of the sectors $b_1\oplus b_1+\xi_1$,
$b_2\oplus b_2+\xi_1$ and $b_3\oplus b_3+\xi_1$.
Three additional 27 and $\overline{27}$ pairs are obtained
from the Neveu-Schwarz $\oplus~\xi_1$ sector.

To construct the model in the orbifold formulation one starts
with the compactification on a torus with nontrivial background
fields \cite{narain}.
The subset of basis vectors,
\beq
\{ 1,S,\xi_1,\xi_2\},
\label{neq4set}
\eeq
generates a toroidally-compactified model with $N=4$ space-time
supersymmetry and ${\rm SO}(12)\times {\rm E}_8\times {\rm E}_8$ gauge
group.
The same model is obtained in the geometric (bosonic) language
by tuning the background fields to the values corresponding to
the SO(12) lattice. The
metric of the six-dimensional compactified
manifold is then the Cartan matrix of SO(12),
while the antisymmetric tensor is given by 
\beq
B_{ij}=\cases{
G_{ij}&;\ $i>j$,\cr
0&;\ $i=j$,\cr
-G_{ij}&;\ $i<j$.\cr}
\label{bso12}
\eeq
When all the radii of the six-dimensional compactified
manifold are fixed at $R_I=\sqrt2$, it is seen that the
left- and right-moving momenta  
$
P^I_{R,L}=[m_i-{1\over2}(B_{ij}{\pm}G_{ij})n_j]{e_i^I}^*
$
reproduce the massless root vectors in the lattice of
SO(12). Here $e^i=\{e_i^I\}$ are six linearly-independent
vielbeins normalized so that $(e_i)^2=2$.
The ${e_i^I}^*$ are dual to the $e_i$, with
$e_i^*\cdot e_j=\delta_{ij}$.

Adding the two basis vectors $b_1$ and $b_2$ to the set
(\ref{neq4set}) corresponds to the ${Z}_2\times {Z}_2$
orbifold model with standard embedding.
Starting from the Narain model with ${\rm SO}(12)\times
{\rm E}_8\times {\rm E}_8$
symmetry~\cite{narain}, and applying the ${Z}_2\times {Z}_2$
twist on the
internal coordinates, reproduces
the spectrum of the free-fermion model
with the six-dimensional basis set
$\{ 1,S,\xi_1,\xi_2,b_1,b_2\}$.
The Euler characteristic of this model is 48 with $h_{11}=27$ and
$h_{21}=3$. I denote the manifold corresponding to this
model as $X_2$.

It is noted that the effect of the additional basis vector $\xi_1$ of eq.
(\ref{vectorx}), is to separate the gauge degrees of freedom, spanned by
the world-sheet fermions $\{{\bar\psi}^{1,\cdots,5},
{\bar\eta}^{1},{\bar\eta}^{2},{\bar\eta}^{3},{\bar\phi}^{1,\cdots,8}\}$,
from the internal compactified degrees of freedom $\{y,\omega\vert
{\bar y},{\bar\omega}\}^{1,\cdots,6}$.
In the realistic free fermionic
models this is achieved by the vector $2\gamma$ \cite{foc}, with
\beq
2\gamma=(0,\cdots,0\vert{\underbrace{1,\cdots,1}_{{\bar\psi^{1,\cdots,5}},
{\bar\eta^{1,2,3}} {\bar\phi}^{1,\cdots,4}} },0,\cdots,0)~,
\label{vector2gamma}
\eeq
which breaks the ${\rm E}_8\times {\rm E}_8$ symmetry to ${\rm
SO}(16)\times
{\rm SO}(16)$.
The ${Z}_2\times {Z}_2$ twist breaks the gauge symmetry to
${\rm SO}(4)^3\times {\rm SO}(10)\times {\rm U}(1)^3\times {\rm SO}(16)$.
The orbifold still yields a model with 24 generations,
eight from each twisted sector,
but now the generations are in the chiral 16 representation
of SO(10), rather than in the 27 of ${\rm E}_6$. The same model can
be realized with the set
$\{ 1,S,\xi_1,\xi_2,b_1,b_2\}$,
by projecting out the $16\oplus{\overline{16}}$
from the $\xi_1$-sector taking
\beq
c{\xi_1\choose \xi_2}\rightarrow -c{\xi_1\choose \xi_2}.
\label{changec}
\eeq
This choice also projects out the massless vector bosons in the
128 of SO(16) in the hidden-sector ${\rm E}_8$ gauge group, thereby
breaking the ${\rm E}_6\times {\rm E}_8$ symmetry to
${\rm SO}(10)\times {\rm U}(1)\times {\rm SO}(16)$.
The freedom in ({\ref{changec}) corresponds to
a discrete torsion in the toroidal orbifold model.
At the level of the $N=4$
Narain model generated by the set (\ref{neq4set}),
we can define two models, ${Z}_+$ and ${Z}_-$, depending on the
sign
of the discrete torsion in eq. (\ref{changec}). The first, say ${Z}_+$,
produces the ${\rm E}_8\times {\rm E}_8$ model, whereas the second, say
${Z}_-$, produces the ${\rm SO}(16)\times {\rm SO}(16)$ model.
The ${Z}_2\times {Z}_2$
twist acts identically in the two models, and their physical
characteristics
differ only due to the discrete torsion eq. (\ref{changec}).

This analysis confirms that the ${Z}_2\times {Z}_2$ orbifold on the
SO(12) Narain lattice is indeed at the core of the
realistic free fermionic models. However, it
differs from the ${Z}_2\times {Z}_2$ orbifold on
$T_2^1\times T_2^2\times T_2^3$, which gives $(h_{11},h_{21})=(51,3)$.
I will denote the manifold of this model as $X_1$.
In \cite{befnq} it was shown that the two models may be connected
by adding a freely acting twist or shift.
Let us first start with the compactified
$T^1_2\times T^2_2\times T^3_2$ torus parameterized by  
three complex coordinates $z_1$, $z_2$ and $z_3$,
with the identification
\beq
z_i=z_i + 1\,, \qquad z_i=z_i+\tau_i \,,
\label{t2cube}
\eeq
where $\tau$ is the complex parameter of each
$T_2$ torus.
With the identification $z_i\rightarrow-z_i$, a single torus
has four fixed points at
\beq
z_i=\{0,{\textstyle{1\over 2}},{\textstyle{1\over 2}}\,\tau,
{\textstyle{1\over 2}} (1+\tau) \}.
\label{fixedtau}
\eeq
With the two ${Z}_2$ twists
\beqn
&& \alpha:(z_1,z_2,z_3)\rightarrow(-z_1,-z_2,~~z_3) \,,
\cr
&&  \beta:(z_1,z_2,z_3)\rightarrow(~~z_1,-z_2,-z_3)\,,
\label{alphabeta}
\eeqn
there are three twisted sectors in this model, $\alpha$,
$\beta$ and $\alpha\beta=\alpha\cdot\beta$, each producing
16 fixed tori, for a total of 48. Adding
to the model generated by the ${Z}_2\times {Z}_2$
twist in (\ref{alphabeta}), the additional shift
\beq
\gamma:(z_1,z_2,z_3)\rightarrow(z_1+{\textstyle{1\over2}},z_2+
{\textstyle{1\over2}},z_3+{\textstyle{1\over2}})
\label{gammashift}
\eeq
produces again fixed tori from the three
twisted sectors $\alpha$, $\beta$ and $\alpha\beta$.
The product of the $\gamma$ shift in (\ref{gammashift})
with any of the twisted sectors does not produce any additional
fixed tori. Therefore, this shift acts freely.
Under the action of the $\gamma$-shift,
the fixed tori from each twisted sector are paired.
Therefore, $\gamma$ reduces
the total number of fixed tori from the twisted sectors   
by a factor of ${2}$,
yielding $(h_{11},h_{21})=(27,3)$. This model therefore
reproduces the data of the ${Z}_2\times {Z}_2$ orbifold
at the free-fermion point in the Narain moduli space.

A comment is in order here in regard to the matching of the 
model that include the shift and the model on the $SO(12)$ lattice.
We noted above that the freely
acting shift (\ref{gammashift}), added to the ${Z}_2\times 
{Z}_2$ orbifold
at a generic point of $T_2^1\times T_2^2\times T_2^3$,
reproduces the data of the ${Z}_2\times {Z}_2$
orbifold acting on the SO(12) lattice.  
This observation 
does not prove, however, that the vacuum which includes the shift
is identical to the free fermionic model. While the 
massless spectrum of the two models may coincide
their massive excitations, in general, may differ.
The matching of the massive spectra is examined by
constructing the partition function of the ${Z}_2\times {Z}_2$
orbifold of an SO(12) lattice, and subsequently
that of the model at a generic point including the
shift. In effect since the action of the ${Z}_2\times {Z}_2$
orbifold in the two cases is identical the problem
reduces to proving the existence of a freely
acting shift that reproduces the partition function of the
SO(12) lattice
at the free fermionic point. Then since the action of 
the shift and the orbifold projections are commuting
it follows that the two ${Z}_2\times {Z}_2$ orbifolds
are identical.

On the compact coordinates there are actually three inequivalent ways
in which the shifts
can act. In the more familiar case, they simply translate a generic point 
by half the
length of the circle. As usual, the presence of windings in string 
theory allows shifts on the T-dual circle, or even asymmetric ones, that 
act both on the circle and on its dual. More concretely, for a circle of
length $2 \pi R$, one can have the following possibilities \cite{vwaaf}:
\beqn
A_1\;:&& X_{\rm L,R} \to X_{\rm L,R} + {\textstyle{1\over 2}} \pi R \,,
\nonumber \\
A_2\;:&& X_{\rm L,R} \to X_{\rm L,R} + {\textstyle{1\over 2}} \left(
\pi R \pm {\pi \alpha ' \over R} \right) \,, 
\nonumber \\
A_3\;:&& X_{\rm L,R} \to X_{\rm L,R} \pm {\textstyle{1\over 2}} {\pi \alpha'
\over R} \,.
\label{a1a2a3}
\eeqn
There is an important difference between these choices: while
$A_1$ and $A_3$ can act consistently on any number of coordinates,
level-matching requires instead that $A_2$ acts on (mod) four real 
coordinates. By studying the respective partition function one 
finds \cite{partitions}
that the shift that reproduces the $SO(12)$ lattice at the
free fermionic point in the moduli space is generated by
the ${Z}_2\times {Z}_2$ shifts
\beqn
g\;: & & (A_2 , A_2 ,0 ) \,,
\nonumber \\
h\;: & & (0, A_2 , A_2 ) \,, \label{gfh}
\eeqn
where each $A_2$ acts on a complex coordinate.
It is then shown that the 
partition function of the SO(12) lattice is reproduced. 
at the self-dual radius, $R_i = \sqrt{\alpha '}$.
On the other hand, the shifts given in Eq. (\ref{gammashift}),
and similarly the analogous freely acting shift given by 
$(A_3,A_3,A_3)$, do not reproduce the partition function
of the $SO(12)$ lattice. 
Therefore, the shift in eq. (\ref{gammashift}) does reproduce
the same massless spectrum and symmetries of the ${Z}_2\times {Z}_2$
at the free fermionic point, but the partition functions of the 
two models differ!
Thus, the free fermionic 
${Z}_2\times {Z}_2$ is realized for a specific form of
the freely acting shift given in eq. (\ref{gfh}).
However, as we saw, all the models that are obtained
from $X_1$ by a freely acting ${Z}_2$-shift have $(h_{11},h_{21})=(27,3)$
and hence are connected by continuous extrapolations.
The study of these shifts in themselves may therefore also yield 
additional information on the vacuum structure of these models
and is worthy of exploration. 

Despite its innocuous appearance the connection between $X_1$ and $X_2$
by a freely acting shift has profound consequences.
First we must realize that any string construction can
only offer a limited glimpse on the structure of string
vacua that possess some realistic characteristics. Thus, the
free fermionic formulation gave rise to three generation
models that were utilized to study issues like Cabbibo
mixing and neutrino masses. On the other hand the free fermionic
formulation is perhaps not the best suited to study
issues that are of a more geometrical character.
We can regard the free fermionic formulation
as heavy duty machinery enabling detailed analysis near a 
single point in moduli space, but obscuring its gross structures.
The geometrical approach on the other
hand provides such a gross overview, but is perhaps less
adequate in extracting detailed properties. 
However, as the precise point where the detailed properties
should be calculated is not yet known, one should regard
the phenomenological success of the free fermionic models
as merely highlighting a particular class of compactified spaces.
These manifolds then possess the overall structure that may
accommodate the detailed Standard Model properties. 
The precise localization of where these properties should be 
calculated, will require further understanding of the 
string dynamics. But, if the assertion that the class of
relevant manifolds has been singled out proves to be correct,
this is already an enormous advance and simplification.

{}From the Standard Model data we may hypothesize
that any realistic string vacuum should possess
at least two ingredients. First, it should contain
three chiral generations, and second, it should
admit their SO(10) embedding. This SO(10) embedding
is not realized in the low energy effective field theory
limit of the string models, but is broken directly at the 
string level. The main phenomenological implication of this
embedding is that the weak-hypercharge has the canonical
GUT embedding. 

It has long been argued that the ${Z}_2\times {Z}_2$ orbifold naturally
gives rise to three chiral generations. The reason being that
it contains three twisted sectors and each of these sectors
produces one chiral generation. The existence of
exactly three twisted sectors arises, essentially, because
we are modding out a three dimensional complex manifold, or
a six dimensional real manifold, by ${Z}_2$ projections,
which preserve the holomorphic three form. Thus, metaphorically
speaking, the reason being that six divided by two equals three. 

However, this argument would hold for any ${Z}_2\times 
{Z}_2$ orbifold
of a six dimensional compactified space, and in particular it 
holds for the $X_1$ manifold. Therefore, we can envision that
this manifold can produce, in principle, models with SO(10) gauge
symmetry, and three chiral generations from the three
twisted sectors. However, the caveat is that this manifold
is simply connected and hence the SO(10) symmetry
cannot be broken by the Hosotani-Wilson symmetry breaking
mechanism \cite{hosotani}.
The consequence of adding the freely acting shift (\ref{gammashift})
is that the new manifold $X_2$, while still admitting
three twisted sectors is not simply connected and hence
allows the breaking of the SO(10) symmetry to one of
its subgroups.

Thus, we can regard the utility of the free fermionic machinery
as singling out a specific class of compactified manifolds. 
In this context the freely acting shift has the crucial
function of connecting between the simply connected covering manifold
to the non-simply connected manifold. Precisely such a construction
has been utilized in \cite{donagi} to construct non-perturbative
vacua of heterotic M-theory. In the next section I discuss these
phenomenological aspects of M--theory. 

\section{M--theory explorations}

The profound new understanding of string theory that
emerged over the past few years means that we can use
any of the perturbative string limits, as well as eleven
dimensional supergravity to probe the properties of 
the fundamental M--theory vacuum.
The pivotal property that this vacuum should preserve
is the $SO(10)$ embedding of the Standard Model spectrum.
Additionally, the underlying compactification should allow
for the breaking of the $SO(10)$ gauge symmetry.
In string theory the prevalent method to break the
$SO(10)$  gauge group is by utilizing Wilson 
line symmetry breaking.

These two properties are not found
in generic string vacua, but are afforded
by the realistic free fermionic models. The free fermionic
models are, however, constructed in the perturbative
heterotic string limit and it is therefore natural to
examine which of their structures is preserved in the 
nonperturbative limit. 
The nonperturbative limit of the heterotic string is conjectured
to be given by the heterotic M--theory limit, or by compactifications
of the Ho\v rava--Witten model \cite{hw} on Calabi--Yau threefolds.
Ho\v rava--Witten theory consists of compactifications of
eleven dimensional supergravity on $S_1/Z_2$.
The orbifold fixed points support ten dimensional supergravity,
with a $E_8$ gauge supermultiplet on each fixed ten dimensional
plane, and the bulk space consists of pure supergravity. 
To construct viable vacua of Ho\v rava--Witten theory
one needs to compactify further to four dimensions
on a Calabi--Yau manifold. 

\begin{figure}[t]
\centerline{\epsfxsize 3.0 truein \epsfbox {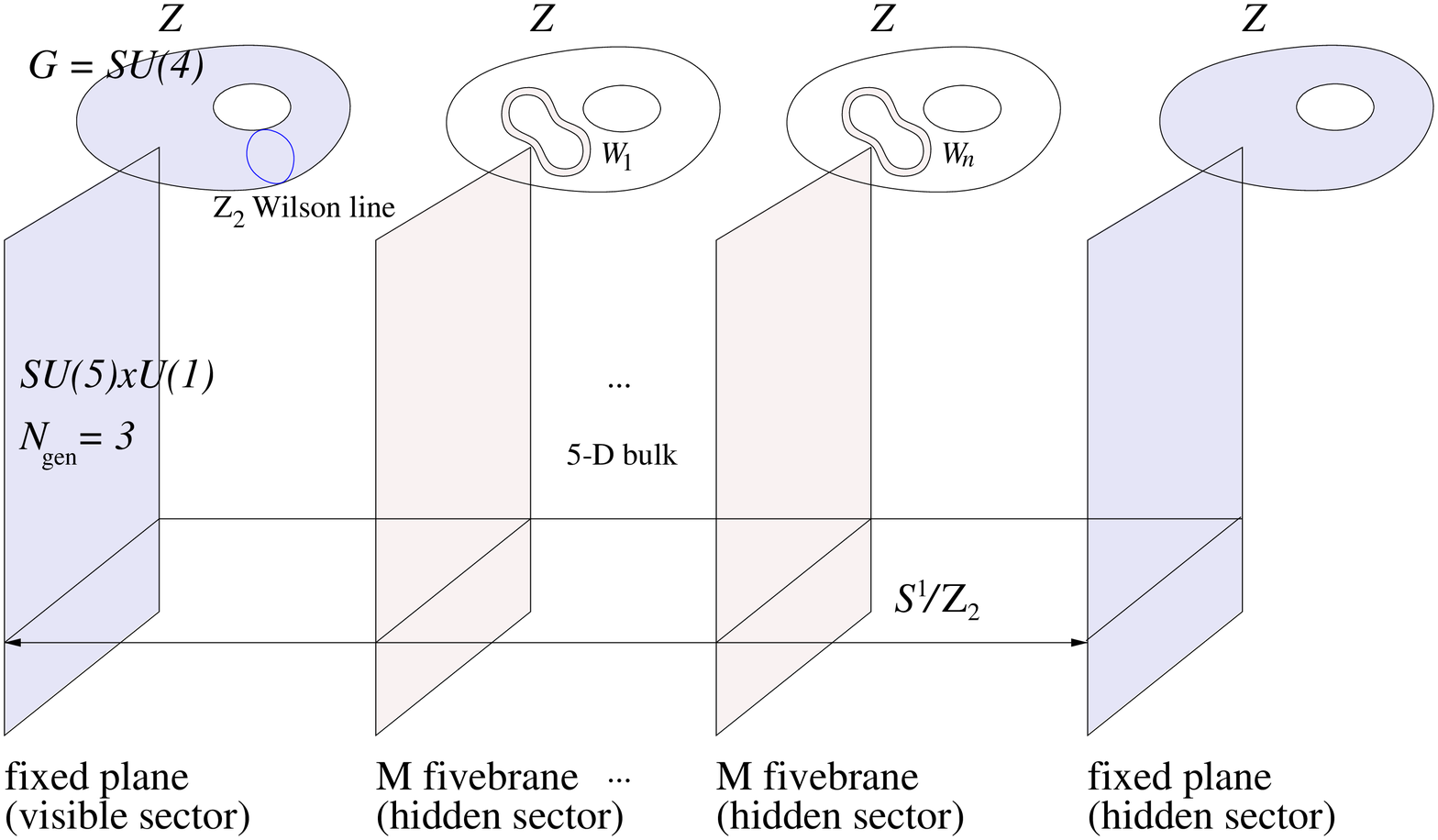}}
\caption{Compactifications of Horava--Witten theory to four dimensions}  
\label{bworld}
\end{figure}

Heterotic M--theory compactifications
to four dimensions have been studied by Donagi {\it et al.}
\cite{donagi}, on manifolds
that do not admit Wilson line breaking and yield $SU(5)$, $SO(10)$
or $E_6$ grand unified gauge groups,
as well as construction of
$SU(5)$ grand unified models that can be broken to the
Standard Model gauge group by Wilson line breaking.
In ref. \cite{fgi}
we extended the work of Donagi \etal to the case of $SO(10)$ models
that allow Wilson line breaking. This entails the modification of
the gauge bundle analysis of ref. \cite{donagi}
from $G=SU(2n+1)$ to $G=SU(2n)$ in the decomposition
of $E_8\supset G\times H$, where $H=SO(10)$ in our case.
The structure of the manifolds constructed by Donagi \etal
correspond to manifolds with fundamental group
${\bf Z}_2$, which is necessary for Wilson line breaking.
In this case the M--theory vacuum allows for $SO(10)$ breaking
by a single Wilson--line.

The construction of Donagi {\it et al.} is very similar
in spirit to the connection between the $X_1$ and $X_2$ 
manifolds discussed in section \ref{z2z2orbifold},
but utilizes a different mathematical language.
As the construction is quite technical, I will not review
it here in detail, but merely give a qualitative overview with
few additional insights from the free fermionic analogy.
The qualitative picture is depicted in figure \ref{bworld}.
In this picture 
The 11--dimensional spacetime $M_{11}$ of M--theory is taken to be
\begin{equation}
M_{11}=M_4\times {S^1\over {\bf Z}_2}\times Z,
\label{spatime}
\end{equation}
where $M_4$ is 4--dimensional Minkowski spacetime, the compact eleventh
dimension
$S^1$ is moded out by the action of ${\bf Z}_2$, and $Z$ is a Calabi--Yau
(complex) 3--fold. There is a semistable holomorphic vector bundle $V_i$,
$i=1,2$ over the 3--fold $Z$ on the orbifold fixed plane at each of the two
fixed
points of the ${\bf Z}_2$--action on $S^1$. The structure group $G_i$ of
$V_i$ is a subgroup of the observable $E_8$.
The new feature of heterotic
M--theory as compared to the perturbative heterotic
string is the existence of 
fivebranes in the vacuum, which wrap holomorphic 2--cycles within
$Z$ and are parallel to the orbifold fixed planes. The fivebranes are
represented by a 4-form cohomology class $[W]$.
The Calabi--Yau 3--fold $Z$, the gauge bundles $V_i$ and the fivebranes are
subject to the cohomological constraint on $Z$
\begin{equation}
c_2(V_1)+c_2(V_2)+[W]=c_2(TZ),
\label{annox}
\end{equation}
where $c_2(V_i)$ is the second Chern class of the $i$--th gauge bundle and
$c_2(TZ)$ is the second Chern class of the holomorphic tangent bundle to $Z$.
Equation (\ref{annox}) above is the anomaly--cancellation condition.

The construction of Donagi \etal is based on generalization of earlier
work of Friedman, Morgan and Witten (FMW) \cite{fmw}.
One starts in this construction
with a Calabi--Yau manifold that admits an elliptic
fibration (denoted $X\equiv X_1$). The Calabi--Yau threefold then
consist of a base, which is a two dimensional complex manifold and
a fiber, which is an elliptic curve.
Friedman, Morgan and Witten derived expressions for
the Chern classes of the gauge and tangent bundles
in terms of the parameters of the base and the fiber
and by utilizing the spectral cover construction.
The derivation of FMW applies to fibrations which admit
a global section. Solutions of different M--theory vacua,
that contain three chiral generations with a GUT gauge group
are then given by writing the effective cohomology classes
of the five branes, $[W]$.
These effective classes are compatible with wrappings of the 
five branes on real 2-cycles of the Calabi--Yau threefolds. 

The existence of a global section in the fibration of $X$
implies that it is simply connected and hence does not admit
GUT Wilson line breaking. Donagi \etal proceed to mod out the
$X$--manifold by a freely acting $Z_2$ involution, $\tau_X$, analogous
to (\ref{gammashift}), yielding a manifold which is not simply
connected (denoted $Z\equiv X_2$).
The key difference between the $X$ and $Z$ manifolds is that
the former admits a global section, whereas the later does not.
The $Z$--manifold admits two sections that are interchanged by
the freely acting involution. This implies that the $Z$ manifold
admits a torus fibration but not an elliptic fibration. 
Additionally, while $\tau_X$ is
freely acting on the Calabi--Yau threefold, it is not freely acting
on the base. Hence, one must blow--up the singular curves on
the base. The involution is guaranteed to be freely acting by
requiring that the intersection of the fixed points of the 
involution with the zeroes of the discriminant of the Weierstrass form
of the fiber is empty. Finally, the expressions for the 
Chern classes of the gauge and tangent bundles and the
effective classes of the five branes are adopted to the 
case of the toroidally fibered Calabi--Yau manifold and
solutions are found for three generation vacua. 

In ref. \cite{fgi} we presented such explicit solutions with 
$SO(10)$ symmetry broken by a Wilson line to $SU(5)\times U(1)$,
which is the flipped $SU(5)$ symmetry breaking pattern \cite{flipped}.
We should note that the construction utilizes a 
single shift and yields $\Pi_1=Z_2$. This implies that
we can break the $SO(10)$ symmetry by a single Wilson line.
Hence, the $SO(10)$ can be broken by a single step to 
$SU(5)\times U(1)$ or $SO(6)\times SO(4)$ but not
directly to the Standard Model gauge group. Similarly, in the
the models built by Donagi \etal a single Wilson line can
be utilized to break the $SU(5)$ symmetry. However, there 
a single breaking step suffices to break $SU(5)$ directly
to the Standard Model gauge group. While the construction of
toroidally fibered manifolds that permit two--step Wilson line breaking is
still underway, we may gain some insight from the free fermionic 
analogs as to what might perhaps be needed. Namely,
in the free fermionic realization of these compactifications
we know explicitly that two, or three, Wilson--line
breakings are possible. Now, we know in fact from the discussion in
section (\ref{z2z2orbifold}) that the free fermionic point is
not realized with the freely acting $\gamma$--shift, Eq. (\ref{gammashift}), 
but rather with the freely acting shift, Eq. (\ref{gfh}).
The task at hand is therefore to implement these more involved 
shifts, which may produce a more complicated $\Pi_1$ structure.

\begin{figure}[t]
\centerline{\epsfxsize 3.0 truein \epsfbox {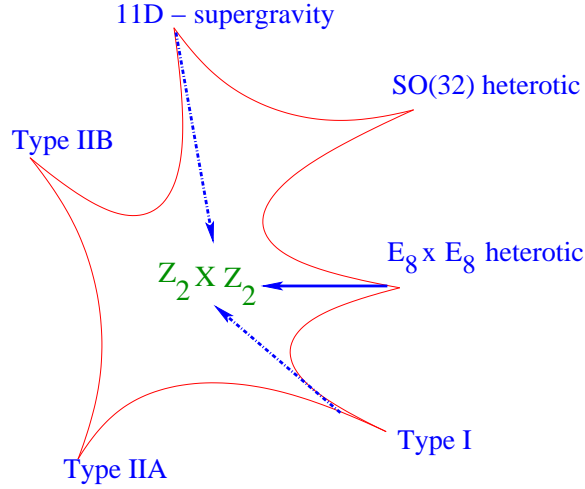}}
\caption{Phenomenological application of M--theory }  
\label{btd02proc1}
\end{figure}

Figure (\ref{btd02proc1}) illustrates qualitatively the approach
to the phenomenological application of M--theory proposed in this
paper. In this view the different perturbative M--theory
limits are used to probe the properties of a specific
class of compactifications. In this respect one may 
regard the free fermionic models as illustrative examples.
Namely, in the heterotic limit this formulation highlighted
the particular class of models that are connected to the
$Z_2\times Z_2$ orbifold. In order to utilize the M--theory
advances to phenomenological purposes, our task then is
to now explore the compactification of the other
perturbative string limits on the same class of
spaces, with the aim of gaining further insight
into their properties. 

\section{Wilsonian matter}\label{uhecr}
The previous sections elaborated on the connection
between the simply and non--simply connected
manifolds, which is obtained by utilizing the freely acting
shifts. This construction
enables the GUT symmetry breaking by Wilson lines. 
In turn, the Wilson line gauge symmetry breaking has far
reaching phenomenological and cosmological implications.
The important consequence is the appearance of the states
in the massless spectrum that do not fit into representations
of the original unbroken GUT gauge group. This phenomena
is peculiar to Wilson--line symmetry breaking and does not arise
in the conventional Higgs symmetry breaking mechanism.
Consequently, field theory GUTs do not give rise, in general,
to such states. Their appearance is a unique
and general consequence of string (M--theory) unification, at least in
the heterotic--string limit. I refer to such states as ``Wilsonian matter''.

In the free fermionic models the Wilsonian matter arises from sectors
that break the $SO(10)$ symmetry, and are produced by combinations of the
basis vectors that extend the NAHE set with the NAHE--set basis vectors.
The $SO(10)$ symmetry is broken to one of its
subgroups, by the assignment of boundary conditions to the
set of complex world--sheet fermions ${\bar\psi}^{1,\cdots,5}$:
\begin{eqnarray}
{b\{{{\bar\psi}_{1\over2}^{1\cdots5}}\}=
\{{1\over2}~{1\over2}~{1\over2}~{1\over2}~
{1\over2}\}
\Rightarrow SO(10)~\rightarrow~SU(5)\times U(1)},~~~\label{so10to64}\\
{b\{{{\bar\psi}_{1\over2}^{1\cdots5}}\}=\{1 ~~1 ~~1 ~~0 ~~0\}
\Rightarrow SO(10)~\rightarrow~SO(6)\times SO(4)}.~\label{so10to51}
\end{eqnarray}
To break the $SO(10)$ symmetry to $SU(3)\times SU(2)\times
U(1)_{B-L}\times U(1)_{T_{3_R}}$
both steps, (\ref{so10to64}) and (\ref{so10to51}),
are used, in two separate basis vectors. The basis vectors $\{\alpha,
\beta,\gamma\}$ that contain the boundary condition assignments eqs.
(\ref{so10to64},\ref{so10to51}) correspond to Wilson lines in the
orbifold construction.

The spectrum of a free fermionic model then contain states from sectors
that do not break the $SO(10)$ symmetry. These typically include the
three generations from the twisted sectors $b_1,~b_2$ and $b_3$;
the Higgs multiplets from the untwisted sector and the sector
$b_1+b_2+\alpha+\beta$; and $SO(10)$ singlets. These states obviously
fit into $SO(10)$ representations. Additionally, the models
contain the states that arise from combinations of the
basis vectors that extend the NAHE set. These states can be
classified according to the pattern of $SO(10)$ symmetry
breaking in each sector. Sectors which break the $SO(10)$
symmetry to $SO(6)\times SO(4)$ or to $SU(5)\times U(1)$
produce states that carry fractional electric charge $\pm1/2$,
which may \cite{revamp,elnfc},
or may not, be confined by a hidden sector non--Abelian
gauge group. Obviously such fractionally charged states should
be either confined, sufficiently massive, or sufficiently rare, 
in order not to conflict with present bounds on the fractionally
charged matter. However, as I discuss further below, their appearance
is perhaps the most dramatic and interesting consequence
of string models. Sectors which break the $SO(10)$ symmetry
directly to the Standard Model gauge group also produce
states with the standard charges under the Standard Model
gauge symmetries, but which carry ``fractional'' charge 
under the $U(1)_{Z^\prime}$ that is embedded in $SO(10)$
and is orthogonal to the Standard Model weak--hypercharge.
Such states can only arise in free fermionic string models
in which the $SO(10)$ symmetry is broken directly to the Standard Model
gauge group. 

The existence of the exotic ``Wilsonian matter'' in the string models
has important phenomenological and cosmological implications. Most 
important is the fact that it provides an intrinsic stringy mechanism
that produces super--heavy meta--stable matter. In the case of the
states that carry fractional electric charge the reason is apparent.
The lightest fractionally charged state
cannot decay into integrally charged matter
and is hence stable. 
More generally, however, we saw in the free fermionic
models that the Standard Model states
fit into $SO(10)$ representations, whereas there exist
exotic Wilsonian states that carry standard charges under the 
Standard Model gauge group, but carry fractional charge
under the $U(1)_{Z^\prime}$. Depending on the charges of the Higgs states
that break the $U(1)_{Z^\prime}$ this case may result in discrete
symmetries that forbid, or highly suppress,
the decay of the exotic states to the Standard Model states. 

\begin{figure}[t]
\centerline{\epsfxsize 3.0 truein \epsfbox {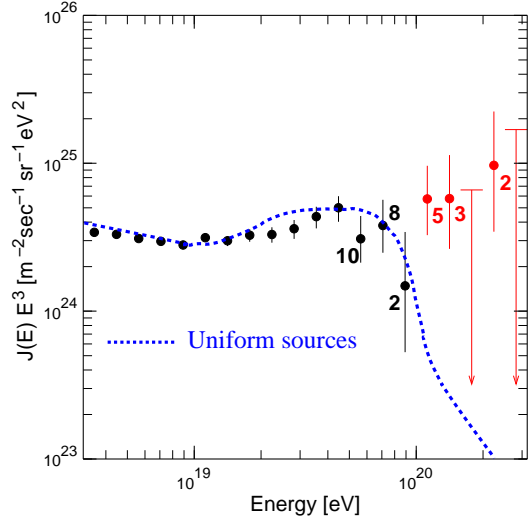}}
\caption{Energy spectrum observed with AGASA \cite{agasa98}}
\label{agasa}
\end{figure}

The appearance of exotic ``Wilsonian matter'' in the string models
is intriguing for another reason.
One of the most fascinating
unexplained experimental observations of recent years is that of
Ultra High Energy Cosmic Rays with energies in excess of
the Greisen--Zatsepin--Kuzmin (GZK) bound \cite{uhecr}.
There are apparently no astrophysical sources
in the local neighborhood that can account for the
events. The shower profile of the
highest energy events is consistent with identification of the
primary particle as a hadron but not as a photon or a neutrino.
The ultrahigh energy events observed in the air shower arrays
have muonic composition indicative of hadrons.
The problem, however, is that the propagation of hadrons
over astrophysical distances is affected by the
existence of the cosmic background radiation, resulting
in the GZK cutoff on the maximum energy of cosmic ray
nucleons $E_{\rm GZK}\le10^{20}\;{\rm eV}$ \cite{gzk}.
Similarly, photons of such high energies have a mean free path of less than
10Mpc due to scattering {}from the cosmic background radiation and
radio photons. Thus, unless the primary is a neutrino,
the sources must be nearby. On the other hand, the primary
cannot be a neutrino because the neutrino interacts very weakly
in the atmosphere. A neutrino primary would imply that the
depths of first scattering would be uniformly distributed
in column density, which is contrary to the observations.
Figure \ref{agasa} shows the number of events as function
of the energy, as observed by the AGASA collaboration \cite{agasa98}.
The dotted curve represent the expected
fall off of events for extra-galactic sources,
because of the GZK cut--off. The AGASA data clearly suggests
the observation of events with energies in excess of the 
GZK cut--off. While this is still a topic of hot debate
among the experimentalists, it is without a doubt one of
the most intriguing pieces of data that emerged in recent
years, and we are well justified in trying to find theoretical
explanations of various sorts. Eventually, it is hoped that
forthcoming observations by the Pierre Auger \cite{auger}
and EUSO \cite{euso} experiments will resolve the experimental issues. 

One of the most intriguing possible solutions
is that the UHECR primaries originate {}from the decay of long--lived
super--heavy relics, with mass of the order of $10^{12-15}\;{\rm GeV}$
\cite{Berezinsky}.
In this case the primaries for the observed UHECR would originate
from decays in our galactic halo, and the GZK bound would not apply.
Furthermore, the profile of the primary UHECR indicates that
the heavy particle should decay into electrically charged
or strongly interacting particles.
{}From the particle physics perspective the meta--stable super--heavy
candidates should possess several properties.
First, there should exist a stabilization mechanism which produces
the super--heavy state with a lifetime of the order of
$
10^{17}s\le \tau_X \le 10^{28}s,~
$
and still allows it to decay and account for the observed UHECR events.
Second, the required mass scale of the meta--stable state
should be of order, $M_X~\sim~10^{12-13}{\rm GeV}.$

Finally, the abundance of the super--heavy relic
should satisfy the relation
$
({\Omega_X/\Omega_{0}})({t_0/\tau_X})\sim5\times10^{-11}\;,~
$
to account for the observed flux of UHECR events.
Here $t_0$ is the age of the universe, $\tau_X$ the lifetime
of the meta--stable state, $\Omega_{0}$ is the critical mass density
and $\Omega_{X}$ is the relic mass density of the meta--stable state.

As discussed above, superstring theory inherently possesses the ingredients
that naturally give rise to super--heavy meta--stable states.
The stabilization
mechanism arises in string theory due to the breaking of the non--Abelian
gauge symmetries by Wilson lines.
The massless spectrum then contains states with fractional electric
charge or ``fractional'' charge under the $SO(10)$ $U(1)_{Z^\prime}$,
which is orthogonal to the weak--hypercharge. 
The lightest fractionally charged state is stable due to electric
charge conservation, and other ``Wilsonian states'' are meta--stable
if their decay to Standard Model states is forbidden by a 
local discrete symmetry \cite{lds}.
In practice it is sufficient to demand that
vevs which break the discrete symmetry are sufficiently small.
The super--heavy
states can then decay via the nonrenormalizable operators
\beq
{{\langle V_1~~~\cdots~~~ V_N\rangle}\over{M_S^{N-3}}}~~~;~~~
M_S\sim10^{17-18}{\rm GeV}
\label{nonreno}
\eeq
which are produced from exchange of heavy string modes.
The lifetime of the meta--stable relic is then given by
\beq
\tau_X~\approx~ {1\over{M_X}}\left(
{{M_S}\over{M_X}}\right)^{2(N-3)}
\label{taux}
\eeq
where $m_X$ is the mass of the meta--stable heavy state,
$M_S$ is the string scale, and $N$ is order of the nonrenormalizable terms. 
Taking $N=8$, $M_S\sim10^{17-18}\;{\rm GeV}$
and $m_x\sim10^{12}\;{\rm GeV}$, one finds that
$\tau_x>10^{7-17}\;{\rm years}$ \cite{elnfc}.

Additionally, string theory may naturally produce mass scales of the
required order, $M_X\approx10^{12-13}{\rm GeV}$. Such mass scales
arise due to the existence of an hidden sector which typically contains
non--Abelian $SU(n)$ or $SO(2n)$ group factors. Thus, the mass scale
of the hidden gauge groups is fixed by the hidden sector gauge
dynamics. Therefore, in the same way that the color $SU(3)_C$
hadronic dynamics are fixed by the boundary conditions at
the Planck scale and the $SU(3)_C$ matter content, the
hidden hadron dynamics are set by the same initial conditions
and by the hidden sector gauge and matter content,
$$M_X~\sim~\Lambda_{\rm hidden}^{\alpha_s,M_S}(N,n_f).$$
Finally, the fact that $M_X\sim10^{12-13}{\rm GeV}$ implies
that the super--heavy relic is not produced in thermal
equilibrium and some other production mechanism is responsible
for generating the abundance of super--heavy relic. This may
arise from gravitational production \cite{ckr} or from inflaton
decay following a period of inflation.

Next I turn to discuss several specific examples. One particular
string theory state that has been proposed as an UHECR candidate
is the 'Crypton', in the context of the flipped $SU(5)$ free fermionic
string model. The `crypton' is a state that carries fractional 
electric charge $\pm1/2$ and transforms under a non--Abelian
hidden gauge group, which in the case of the flipped 
$SU(5)$ ``revamped''string model is $SU(4)$. 
The fractionally charged
states are confined and produce integrally charged hadrons
of the hidden sector gauge group, which have been named ``tetrons''.
The lightest hidden hadron
is expected to be neutral with the heavier modes split
due to their electromagnetic interactions. A priori, therefore,
the `crypton' is an appealing CDM candidate, which is meta--stable
because of the fractional electric charge of the constituent `quarks'.
This implies that the decay of the exotic hadrons
can be generated only by highly suppressed non--renormalizable
operators (\ref{nonreno}). Effectively, therefore, the events that generate
the UHECR are produced by annihilation of the `cryptons'
in the confining hidden hadrons. Moreover, the mass scale
of the hidden hadrons is fixed by the hidden sector gauge
dynamics, and may naturally be of the order $10^{12-13}{\rm GeV}$.
Therefore, in the same way that the color $SU(3)_C$
hadronic dynamics are fixed by the boundary conditions at
the Planck scale and the $SU(3)_C$ matter content, the 
hidden hadron dynamics are set by the same initial conditions
and by the hidden sector matter content.
Depending on the order $N$ of the nonrenormalizable tetron--decay
mediating operators, the lifetime from Eq. (\ref{taux})
may be in the appropriate range to account for the flux of observed
UHECR events above the GZK cutoff \cite{elnfc}. However, in addition to
the lightest neutral bound states there exist in this model
also long lived meta--stable charged bound states, whose abundance
is comparable to that of the neutral states \cite{cfp}.
The reason is that the cryptons, which carry electric
charge $\pm1/2$, are singlets of $SU(2)_L$. Therefore, the charged
tetrons cannot decay to the neutral tetron by emitting a
light $W^\pm$ or a light Higgs. Consequently, the charged tetrons
can only decay by emitting a massive string state. Effectively,
therefore, the only way for the charged tetrons to decay to the
neutral one is by the same higher dimensional operators that
govern the decay of the neutral tetrons.
The lifetime of the charged and neutral tetrons are
consequently of similar order. 
Constraints on the abundance of stable charged heavy matter
then places an additional constraint on the lifetime of
this form of UHECR candidates. One finds \cite{cfp}
that for a tetron mass $\sim 10^{12}{\rm GeV}$
the tetron density is at most
$\Omega_T\le 10^{-6} \Omega_{CDM}\;.$ However, the
tetron can still account for the observed UHECR events,
provided that $\tau_T\le 2\times10^4t_U$, where
$t_U$ is the age of the universe. Additionally,
since the crypton carry fractional $U(1)_Y$ charge
$\pm1/2$ and transform as $4$ and $\bar4$ of the 
hidden $SU(4)$ gauge group, it affects the evolution
of the weak--hypercharge gauge coupling, but not that of
$\alpha_2$ or $\alpha_3$. Consequently, coupling unification
necessitates the introduction of additional matter to
achieve unification.

In addition to the fractionally charged states, the free fermionic
standard--like models contain states, which arise from
$SU(3)\times SU(2)\times U(1)^2$ type sectors, and
carry the regular charges under the Standard Model,
but carry ``fractional'' charges under the $U(1)_{Z^\prime}\in SO(10)$
symmetry. These states can be color
triplets, electroweak doublets, or Standard Model singlets
and may be good dark matter candidates \cite{ccf}.
The meta--stability of this type of states arises because
of their fractional $U(1)_{Z^\prime}$ charge. Namely, the fact that
the Standard Model states possess the $SO(10)$ embedding,
implies that there exist a discrete symmetry which protects
the exotic matter from decaying into the lighter Standard Model states.
We must additionally insure that the $U(1)_{Z^\prime}$ symmetry breaking
VEVs, break the discrete symmetry sufficiently
weakly. The uniton is such a color triplet that has been motivated
to exist at an intermediate energy scale due to its possible
role in facilitating heterotic--string gauge coupling unification.
It forms bound states with ordinary down and up quarks. The mass
of the uniton is generated from nonrenormalizable terms and
can be of order $10^{12-13}{\rm GeV}$, as required to explain
the UHECR events. Additionally, if the uniton is to contribute
substantially to the dark matter, the lightest bound state
must be neutral and the heavier charged states must be unstable.
However, contrary to the case of the fractionally charged states,
the uniton charged bound states can decay through $W^\pm$ radiation of
the ordinary quark with which it binds. Obviously, the uniton
also affects the evolution of the Standard Model gauge
couplings. However, a uniton state at intermediate mass scale
was motivated precisely because of its effect on the
Renormalization Group Equations. Therefore, intermediate
mass scale uniton fulfill the dual role of facilitating
heterotic--string gauge coupling unification, and at the 
same time providing a dark matter candidate and a meta--stable
super--heavy state that may explain the UHECR events.

Lastly, the free fermionic
Standard--like models contain Standard Model singlets that carry
fractional $U(1)_{Z^\prime}\in SO(10)$ charge. Such states
may be semi--stable provided that the discrete symmetry, which
suppresses their decay modes, is
broken sufficiently weakly. Similar to the states
with fractional electric charge, they may transform under a
hidden sector non--Abelian gauge group and their
mass scale may therefore be fixed by the confining
hidden sector dynamics. They do not affect the evolution of
the Standard Model gauge couplings, and being neutral,
they provide ideal dark matter and UHECR candidates.

\section{Blessings in the sky}

We observed in the previous section that
superstring models provide a variety of candidates, with differing
properties, that may account for the observed UHECR events.
The phenomenological challenge is to develop
the tools that will discern between the different candidates,
by confronting their intrinsic properties
with the observed spectrum of the cosmic ray showers.
The UHECR data, however, opens up new probes to the GUT
and string scale physics. The point is that in the analysis
of the decay products of the meta--stable states one must
extrapolate measured parameters from the low scale, at which
they are measured, to the high--scale of the hypothesized meta--stable
state. In this extrapolation, which covers more than 10 orders of magnitude
in energy scales, one must make some judicious assumptions
in regard to the particle content. Thus one may hope that the
extrapolation itself will enable to differentiate between
different assumptions in regard to the physics in the extrapolation range.
This methodology is very similar to that employed successfully in the
case of gauge coupling unification in supersymmetric versus
non--supersymmetric cases \cite{gcumssm}.
There the motivation for the extrapolation
arises from the hypothesis of unification and one can show that it is
consistent only if one includes the supersymmetric spectrum.
Similarly, in the case of the UHECR events, the motivation for
the high scale are the events themselves and the possibility
to explain them with the super--heavy meta--stable matter. The
extrapolated parameters are the QCD fragmentation functions and
similarly one must include in the evolution whatever physics
is assumed to exist in the desert. In ref. \cite{cf}
supersymmetric fragmentation functions were developed for this
purpose. Furthermore, such functions may also be used in the
analysis of the cosmic rays showers, which arise from the
collision of the primaries with the atmosphere nuclei at a center of mass
energies of order $\sim100{\rm TeV}$. 

\begin{figure}[t]
\centerline{\epsfxsize 3.0 truein \epsfbox {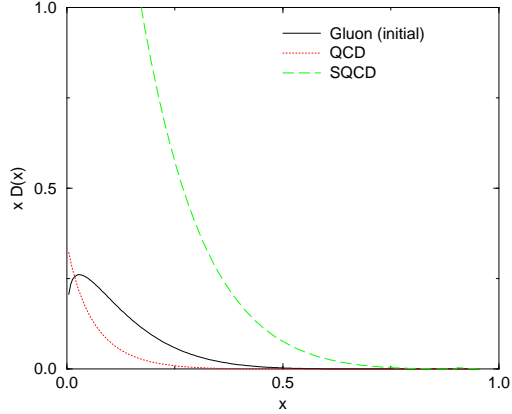}}
\caption{
The gluon fragmentation function $x D_{g}^{p,\bar{p}}(x,Q^2)$
at the lowest scale $Q_0=10$ GeV,
and its evolved QCD (regular) and SQCD/QCD evolutions 
with $Q_f=$$10^3$ GeV.
The SUSY fragmentation scale is chosen to be $200$ GeV.}  
\label{gluonx}
\end{figure}

As an illustration of the procedure, consider the decay 
of a hypothetical 1 TeV massive particle into supersymmetric partons. 
The decay can proceed, for instance, through a regular $q\bar{q}$ 
channel, and a shower is developed starting from the quark pair.
The $N=1$ DGLAP 
equation describes in the leading log approximation the evolution 
of the shower which accompanies the pair.
We are interested in studying the 
impact of the supersymmetric spectrum on the fragmentation. 
In our runs we have chosen the initial set of Ref.~\cite{kkp}.          
We parameterize the fragmentation functions as
\begin{equation}
\label{temp}
D(x,\mu^2)=Nx^\alpha(1-x)^\beta\left(1+\frac{\gamma}{x}\right)
\end{equation}

Typical fragmentation functions in QCD involve final states with 
$p$, $\bar{p}$, $\pi^{\pm},\pi^{0}$ and kaons $k^{\pm}$.
We have chosen an initial evolution scale of $10$ GeV and
varied both the common SUSY mass scale, $Q_i$,
and the final evolution scale, $Q_f$. 
In general the effects of supersymmetric evolution 
are small within the range described by the factorization scales $Q_f$ and 
$Q_i$ ($Q_f=10^3$ GeV, $Q_i=200$ GeV). 

The situation appears to be completely different 
for the gluon fragmentation functions.
In figure \ref{gluonx} the evolution using
the supersymmetric and non--supersymmetric
fragmentation functions is displayed.
The regular and the SQCD evolved fragmentation functions
differ largely in the diffractive region. This
will show up in the spectrum of the primary protons 
if the decaying state has a supersymmetric content. 
As we raise the final evolution scale 
we start seeing more pronounced differences between regular 
and supersymmetric distributions.
Comparison of the squark fragmentation functions for all the flavors and 
the one of the gluino \cite{cf} shows that the scharm distribution
grows slightly faster than the scalar distributions. The gluino
fragmentation function is still the fastest growing at 
small--x values \cite{cf}.

The second important scale appearing in the
analysis of UHECR concerns the interaction of the primary protons
with the nuclei in the atmosphere. These interactions, estimated to be
in the several TeV's, may require a supersymmetric analysis.
The supersymmetric evolution of the parton distributions 
of gluons, squarks and gluinos reveals
that the small-x rapid growth of the gluino distribution 
is visible and points toward an interesting effect 
in the hadronic cross section of the primaries \cite{cf}.
Most exciting, however, is perhaps
the fact that the forthcoming Pierre Auger and EUSO experiments
will explore precisely the physics of the UHECR above the GZK cutoff!
The hypothesized meta--stable super--heavy string relics
may then serve as experimental probes of the string physics,
provided that we are able to develop the phenomenological tools
to decipher their predicted properties, such as their fractional
electric or $U(1)_{Z^\prime}$ charge! Initial efforts in this
direction were recently reported in ref. \cite{ccgm}.

\section{Fundamental principles}

Over the past few years important progress has been achieved
in the basic understanding of string theories. Under the umbrella
of M--theory they are seen to be limits of one fundamental theory.
As such M--theory provides the most advanced framework to date
to study the unification of the gauge and gravitational interactions.
This paradigm led to explicit and detailed models, in
the various perturbative limits, and the methodology for
confronting those models with experimental data is 
under intense development. However, should this be the end
of the road for M-theory? The answer is obviously not! 
The story of M--phenomenology, as it pertains to observable
phenomena, will not be complete without a rigorous formulation
that embarks from well posited basic physics axioms, \`a la
classical, special and general relativity.

It may however be that such a basic formulation
requires a revision of the basic formalism
of quantum mechanics and quantum field theories. 
Despite their unprecedented success in the context
of the Standard Models, there may exist a fundamental
discrepancy between the utilization of the vacuum
in quantum field theories, versus what may be needed
in the eventual formulation of quantum gravity.
In this context, in collaboration with Marco Matone
from the University of Padova, we presented a formulation
of quantum mechanics starting from an equivalence postulate
\cite{fm2,bfm}.
I propose that our approach may serve as a starting point
for the rigorous formulation of quantum gravity.
I recap here briefly the two key ingredients of the
formalism whose generalization may yield such a formulation. 

The equivalence postulate states
that all physical systems labeled by the function
${\cal W}(q)=V(q)-E$, can be connected by a coordinate
transformation, $q^a\rightarrow q^b=q^b(q^a)$, defined
by ${\cal S}_0^b(q^b)={\cal S}_0^a(q^a)$.
This postulate implies that there
always exist a coordinate transformation connecting
any state to the state ${\cal W}^0(q^0)=0$.
Classically the state ${\cal W}^0(q^0)=0$ is a fixed point
in the space of allowed solutions and therefore the
equivalence postulate cannot be implemented consistently.
Consistency of the
equivalence postulate implies the modification of CM,
which is analyzed by a adding a still unknown function
$Q$ to the Classical Hamilton--Jacobi Equation.
Consistency of the equivalence postulate fixes the
transformation properties for ${\cal W}(q)$,
$
{\cal W}^v(q^v)=
 \left(\partial_{q^v}q^a\right)^2{\cal W}^a(q^a)+(q^a;q^v),
$
and for $Q(q)$,
$
 Q^v(q^v)=\left(\partial_{q^v}q^a\right)^2Q^a(q^a)-(q^a;q^v),
$
which fixes the cocycle condition for the inhomogeneous term
\beq
(q^a;q^c)=\left(\partial_{q^c}q^b\right)^2[(q^a;q^b)-(q^c;q^b)].
\label{cocycle}
\eeq
The cocycle condition is the first key ingredient in the
formalism. It is invariant under M\"obius transformations
that fix the functional form of the inhomogeneous term.
The cocycle condition is generalized to higher, Euclidean or
Minkowski, dimensions \cite{bfm},
where the Jacobian of the coordinate transformation extends
to the ratio of momenta in the transformed and original systems.
A second key ingredient in the formalism is the identity
\beq
({\partial_q{\cal S}_0})^2=
\hbar^2/2\left(\{\exp(i2{\cal S}_0/\hbar,q)\}-\{{\cal S}_0,q\}\right)
\label{schwarzianidentity}
\eeq
where $\{,\}$ denotes the Schwarzian derivative, and
${\cal S}_0$ is solution of the Quantum Stationary
Hamilton--Jacobi Equation. The fundamental characteristic
of quantum mechanics in this approach is that ${\cal S}_0\ne Aq+B$
always. The equivalence postulate may also shed light
on the quantum origin of mass. The generalization of
the identity (\ref{schwarzianidentity})
to the relativistic case with a vector potential is,
$$
\alpha^2(\partial{\cal S}-eA)^2={D^2(Re^{\alpha{\cal S}})/
(Re^{\alpha{\cal S}})}-{\Box R/ R}-({\alpha/ R^2})
\partial\cdot(R^2(\partial {\cal S}-eA)),
$$
where $\alpha=i/\hbar$, $D$ is a covariant derivative,
and $\partial\cdot(R^2(\partial{\cal S}-eA))=0$ is a
continuity condition. The $D^2(Re^{\alpha{\cal S}})/
(Re^{\alpha{\cal S}})$ term is associated with the
Klein--Gordon equation. In this case ${\cal W}(q)=1/2 m c^2$.
{}From the equivalence postulate it follows that masses
of elementary particles arise from the inhomogeneous
term in the transformation of the ${\cal W}^0(q^0)\equiv0$
state, {\it i.e.}
$
1/2 m c^2= (q^0;q).
$
{}From this perspective we may speculate that scalar particles
and symmetry breaking represent a particular realization
of the geometrical transformation $q^0\rightarrow q$.
Obviously, this interpretation offers new possibilities
to understand how particle properties are generated from
the vacuum. Consistency of the equivalence postulate
also implies tunneling and energy quantization for bound states
without assuming the probability interpretation of the
wave function, and may therefore offer new perspective
on the origin of the Hilbert space structure \cite{fm2}.

The equivalence postulate approach offers a rigorous formalism
for quantum mechanics. While its formulation is still in its
infancy, we note that the main phenomenological characteristics
of quantum mechanics emerge from its consistent implementation.
The structure of the formalism is mathematically elegant and rich.
The generalization of the two basic
ingredients, Eq. (\ref{cocycle}) and Eq. (\ref{schwarzianidentity}),
may eventually yield a rigorous formulation of quantum gravity. 

\section{Conclusions}

Twentieth century particle physics began with the discovery of the
the electron by Thompson in 1897. By its end the gauge and matter
sectors of the Standard Model have been firmly established.
Meanwhile, experimental particle physics evolved
from the cathode ray tubes used by Thompson to present day
Mega--colliders. In this process, the societies hosting these experiments
have been transformed irreversibly, and future impact 
can be nothing but a speculation. 

The Standard Model opened the door to the possibility of realizing
Einstein's dream of finding a unified mathematical framework
for the known fundamental matter and interactions.
We should not disregard the fact that experimental
elucidation of the Higgs mechanism is determinantal
to fulfilling this program, and in its lack basic understanding
of the origin of mass and the nature of the vacuum
is not possible. 

The available experimental data strongly
indicates the embedding of the Standard Model spectrum
in grand unified theories, most
elegantly in $SO(10)$.
The logarithmic running of the Standard Model
gauge and matter parameters is confirmed by experiments,
and is in qualitative agreement with
high scale unification.
Understanding of the Standard Model flavor structure
requires unification of the gauge and gravitational
interactions. String theories provide the contemporary
tools to study this unification.

The development of M--theory transformed our understanding
of string theories, which are seen to be limits of a single more
fundamental theory. This gives new context to attempts to
connect between string theory and experimental data. Indeed
numerous studies are underway to construct
viable string models by using branes and open string
constructions \cite{branemodels,typeImodels}. The approach reviewed
in this paper is slightly different. In this view
the different string limits can only
probe some properties of the true nonperturbative vacuum.
Thus, in the heterotic limit the grand unification
structures underlying the Standard Model can be seen.
On the other hand, the heterotic string is expanded
in the zero coupling limit and finite coupling requires
moving away from the perturbative heterotic--string limit.
Thus, dilaton stabilization might be better addressed
in the type I limit.
In this respect the fundamental question remains, the
selection of the compactified space, out of the myriad
of the possibilities, that may correspond to our world.
We may regard the string duality picture as allowing
us to probe the properties of specific compactifications
by compactifying the different limits on the same class of
manifolds.

In the heterotic limit the non--Abelian GUT symmetries
are broken by Wilson lines, which gives rise to exotic matter.
In turn, this exotic matter provides several candidates for
dark matter and UHECR events, with differing characteristics.
One must therefore develop the
tools to decipher their phenomenological properties.
Most exciting in this regard is the fact that the Pierre Auger
and EUSO experiments are designed to study the physics of UHECR
events beyond the GZK cut-off, and some surprises may lie in store!

\section*{Acknowledgments}

I would like to thank
Carlo Angelantonj, Claudio Coriano, Richard Garavuso, Jose Isidro,
Marco Matone and Michael Pl\" umacher for collaboration
and discussions, and the CERN theory group for hospitality.
This work is supported in part by PPARC.



\section*{References}
\begin{thebibliography}{9}
\normalsize
\bibitem{Mtheoryreviews} Townsend P K hep-th/9612121;
                         Sen A hep-th/9802051;                  
                         Duff M J hep-th/9805177;
                         Ovrut B hep-th/0201032.

\bibitem{gg}      Georgi H and Glashow S 1974 \PRL {\bf32} {438}.

\bibitem{so10}  Georgi H in {\it Particles and Fields--1974},
                    ed. C.E. Carlson. 1975, New York, AIP Press;
        Fritzsch H and Minkowski P 1975 Annals Phys. {\bf93} 193.

\bibitem{lowerscale} Witten E \NPB{471}{1996}{135};
                        Lykken J \PRD{54}{1996}{3693};
                        Antoniadis I \etal \PLB{436}{1998}{257}; 
                        Dienes K R \etal \PLB{436}{1998}{55}.

\bibitem{gqw} Georgi H \etal 1974 \PRL {\bf33} {451}.
\bibitem{gcumssm} Dimopoulos \etal \PRD{24}{1981}{1681};
                  Ellis J \etal \PLB{260}{1991}{131};
                  Langacker P and Luo M \PRD{44}{1991}{817};
                  Amaldi U\etal \PLB{260}{1991}{447}.

\bibitem{mbmtau} Chanowitz M \etal \NPB{128}{77}{506};
                 Buras A J \etal \NPB{135}{78}{66}.

\bibitem{suthree} Candelas P \etal
                                        \NPB{258}{1985}{46};
                Greene B \etal
                                        \NPB{292}{1987}{606};
                Arnowitt R and  Nath P 1989 \PRL {\bf62} {2225};
\bibitem{twozero} Witten E \NPB{268}{1986}{79}; 
                  Distler J and Kachru S \NPB{413}{1994}{213}.
\bibitem{dhvw} Dixon L \etal \NPB{274}{1986}{285}.
\bibitem{narain} Narain K S \PLB{169}{1986}{41};
                 Narain K S \etal \NPB{279}{1987}{369}.
\bibitem{zthree} Font A \etal \NPB{331}{1990}{421};
                Casas J A \etal\NPB{317}{1989}{171}.
\bibitem{giedt} Giedt J 2002 Annals Phys. {\bf297} 67.
\bibitem{FFF} Kawai K \etal \NPB{288}{1987}{1};
                Antoniadis I \etal \NPB{289}{1987}{87}.
\bibitem{gepner}  Gepner D \NPB{290}{1987}{10};
\NPB{290}{1988}{757}.
\bibitem{foc} \AEF \PLB{326}{1994}{62}.
\bibitem{btd9799} \AEF hep-ph/9707311; hep-th/9910042.
\bibitem{KLN} Dixon L \etal \NPB{282}{1987}{13};
              Kalara S \etal \NPB{353}{1991}{650}.
\bibitem{dsw}
         Dine M \etal \NPB{289}{1987}{589};
         Atick J J \etal \NPB{292}{1987}{109};
         Cecotti S \etal \IJMP{2}{1987}{1839}.
\bibitem{nahe} Faraggi A E and Nanopoulos D V \PRD{48}{1993}{3288};
               Faraggi A E hep-th/9511093; hep-th/9708112.

\bibitem{revamp} {Antoniadis I \etal \PLB{231}{1989}{65}.}
\bibitem{patisalamstrings} Antoniadis I \etal
                                        \PLB{245}{1990}{161};
                Leontaris G K and Rizos J \NPB{554}{1999}{3}.
\bibitem{fny} \AEF \etal \NPB{335}{1990}{347};
		Faraggi A E \PRD{46}{1992}{3204}.
\bibitem{eu} \AEF \PLB{278}{1992}{131}; \NPB{387}{1992}{239}.
\bibitem{top} \AEF \PLB{274}{1992}{47}; \PLB{377}{1995}{43};
                \NPB{487}{1996}{55}.

\bibitem{cfn} Cleaver G B \etal \PLB{455}{1999}{135};
                                \IJMP{16}{2001}{425};
                                \NPB{593}{2001}{471};
                                \MODA{15}{2000}{1191};
                                \IJMP{16}{2001}{3565};
                                \NPB{620}{2002}{259}.
\bibitem{lrsstringmodels}
                Cleaver G B \etal \PRD{63}{2001}{066001};
                                   \PRD{65}{2002}{106003}.
\bibitem{ccf}   Chang S \etal   \PLB{397}{1997}{76};
                                \NPB{477}{1996}{65}; 
              Elwood J and \AEF \NPB{512}{1998}{42}.
\bibitem{otherrsm} Chaudhuri S \etal \NPB{469}{1996}{357};
\bibitem{penn} Cleaver G B \etal \NPB{525}{1998}{3};
                                \NPB{545}{1998}{47};
                                \PRD{59}{1999}{055005};
                                \PRD{59}{1999}{115003}.
\bibitem{review} For review see {\it e.g.}:
                        Lykken J hep-ph/9511456;
                        Lopez J L hep-ph/9601208;
                        Faraggi A E hep-ph/9404210.
\bibitem{cdfd0} Abe F \etal 1995 \PRL {\bf74} {2626};
                Abachi S \etal 1995 \PRL {\bf74} {2632}.
\bibitem{NRT} \AEF \NPB{403}{1993}{101};\NPB{407}{1993}{57}.
\bibitem{CKM} Antoniadis I \etal \PLB{278}{1992}{257};
\AEF and Halyo E \PLB{307}{1993}{305}; \NPB{416}{1994}{63};
Ellis J \etal \PLB{425}{1998}{86}.
\bibitem{seesaw} Antoniadis I \etal \PLB{279}{1992}{281};
             Faraggi A E and Halyo E \PLB{307}{1993}{311};
             Faraggi A E and Pati J C \PLB{400}{1997}{314}. 
\bibitem{gcu} Antoniadis I \etal \PLB{268}{1991}{188}; 
              Antoniadis I \etal \PLB{272}{1991}{31};
              Faraggi A E \PLB{302}{1993}{202}; 
              Dienes K R and Faraggi A E 1995 \PRL {\bf75} {2646};
                                           \NPB{457}{1995}{409}.             
\bibitem{ps} Faraggi A E \NPB{428}{1994}{111};
                         \PLB{499}{2001}{147};
                         \PLB{520}{2001}{337};
             Ellis J \etal \PLB{419}{1998}{123}.
\bibitem{fp2} Antoniadis I \etal \PLB{241}{1990}{24}; 
              \AEF and Halyo E \IJMP{11}{1996}{2357};
              Faraggi A E and Pati J C \NPB{526}{1998}{21};
              Faraggi A E and Vives O hep-ph/0203061.
\bibitem{zp} Costa G \etal \NPB{297}{1988}{244};
             Hewett J and Rizzo T \PRT{183}{1989}{193};
             Faraggi A E and Nanopoulos D V \MODA{6}{1991}{61};
             Cvetic M and Langacker P hep-ph/9707451.
\bibitem{dedes} Dedes A and Faraggi A E \PRD{62}{2000}{016010}.
\bibitem{rp} Faraggi A E \PLB{398}{1997}{95}.
\bibitem{custodial} \AEF \PLB{339}{1994}{223}.
\bibitem{thor} \AEF and Thormeier M \NPB{624}{2002}{163}. 
\bibitem{jogesh} Pati J C, \PLB{388}{1996}{532}.
\bibitem{dienes} Dienes K R and March--Russell J \NPB{479}{1996}{113}.
\bibitem{fccp} Wen X G and Witten E, \NPB{261}{1985}{651};
               Schellekens A, \PLB{237}{1990}{363};

\bibitem{elnfc}  Ellis J \etal
                        \PLB{247}{1990}{257};
		Ellis J \etal \NPB{373}{1992}{399};
                Benakli K \etal
                                        \PRD{59}{1999}{047301};
                Sarkar S and Toldra R \NPB{621}{2002}{495}.

\bibitem{fop} \AEF, Olive K A and Pospelov M \APJ{13}{2000}{31}.
\bibitem{cfp} Corian\`{o} C, \AEF~and Plumacher M, \NPB{614}{2001}{233}.

\bibitem{befnq} Berglund P \etal \PLB{433}{1998}{269}; \IJMP{15}{2002}{1345}.

\bibitem{vwaaf} Vafa C and Witten E 1996
	{\it Nucl.Phys.Proc.Suppl.} {\bf46} 225;
	Angelantonj C \etal \NPB{555}{1999}{116}.
\bibitem{partitions} \AEF hep-th/0206165, to appear in PLB.
\bibitem{hosotani} Hosotani Y, \PLB{126}{83}{309}; \PLB{129}{83}{193}. 

\bibitem{donagi} Donagi R \etal hep-th/9901009; hep-th/9912208;
{\it Class. Quant. Grav.} {\bf 17} (2000) 1049.

\bibitem{hw} Ho\v rava H and Witten E \NPB{460}{1996}{506};
\NPB{475}{1996}{94}.

\bibitem{fgi}\AEF \etal hep-th/0204080, to appear in NPB. 

\bibitem{fmw} Friedman R \etal 1997 {\it Comm. Math. Phys.} {\bf 187} 679.

\bibitem{flipped} Barr S M \PLB{112}{1982}{219};
                  Derendinger J P \etal
                             \PLB{139}{1984}{170};
                  Antoniadis I \etal
                             \PLB{194}{1987}{231}.
\bibitem{uhecr} Hayashida N \etal 1994 \PRL {\bf73} {3491};
                Bird D J \etal \APJ{424}{1994}{491}.

\bibitem{gzk} Greisen K 1996 \PRL {\bf16} {748};
              Zatsepin G T and Kuzmin V A 1996
         {\it Pisma Zh. Eksp. Theor. Fiz.} {\bf 4} 114.

\bibitem{agasa98} Takeda M \etal 1998 \PRL {\bf81} 1163.


\bibitem{auger}The Auger Collaboration 2002
{\it Nucl.Phys.Proc.Suppl.} {\bf110} 487.
\bibitem{euso} Catalano O \nuvc{24}{2001}{445}.


\bibitem{Berezinsky}
Berezinsky V, Kachelriess{} M and Vilenkin A 1997 \PRL {\bf79} {4302};
Kuzmin V A and Rubakov V A, 1998
        {\it Phys. Atom. Nucl.} {\bf 61} 1028.

\bibitem{lds} Krauss L M and Wilczek F 1989 \PRL {\bf62} {1221};
              Faraggi A E \PLB{398}{1997}{88}.

\bibitem{ckr}
Chung D J \etal \PRD{59}{1999}{023501}.


\bibitem{cf}  Corian\`{o} C and \AEF \PRD{65}{2002}{075001};
hep-ph/0106326; hep-ph/0201129.


\bibitem{kkp}Kniehl B A \etal \NPB{582}{2000}{514}.

\bibitem{ccgm} Cafarella A \etal hep-ph/0208023.

\bibitem{fm2} \AEF and Matone M \PLB{450}{1999}{34}; \PLB{437}{1998}{369};
\PLA{249}{1998}{180}; \PLB{445}{1998}{77}; \PLB{445}{1999}{357};
\IJMP{15}{2000}{1869}; \AEF hep-th/0003156; Matone M hep-th/0005274.
\bibitem{bfm} Bertoldi G \etal hep-th/9909201.

\bibitem{branemodels} Aldazabal G \etal \JHEP{0002}{2000}{015};
\JHEP{0102}{2001}{047};
Antoniadis I \etal \PLB{464}{1999}{38}; \PLB{486}{2000}{186};
Bailin D \etal \PLB{502}{2001}{209}; \PLB{530}{2002}{202};
Blumenhagen \etal \JHEP{0010}{2000}{006}; \NPB{616}{2001}{3};
\JHEP{0207}{2002}{026};
Kakushadze \etal \NPB{533}{1998}{25};
Cvetic M \etal \NPB{615}{2001}{3}; hep-th/0206115;
Ibanez L E \etal \NPB{542}{1999}{112}; \JHEP{0111}{2001}{002};
Ellis J \etal hep-th/0206087; Kokorelis C hep-th/0207234.

\bibitem{typeImodels} For review and references see {\it e.g.}:
C. Angelantonj and A. Sagnotti, hep-th/0204089.

\end{thebibliography}
\end{document}